\documentclass[twocolumn,a4paper]{article}

\usepackage{amsmath,amssymb,amsfonts,mathrsfs}
\usepackage{epsfig,epstopdf}
\usepackage{graphicx,graphics}
\usepackage{color}
\usepackage{float}
\usepackage{ifpdf}
\usepackage[colorlinks=true]{hyperref}
\usepackage{braket}

\usepackage{mathtools,lipsum, nccmath,etoolbox}

\usepackage{cuted}
\def\mbb{\mathbb}
\newtheorem{theorem}{Theorem}
\newtheorem{corollary}{Corollary}

\newtheorem{example}{Example}
\newtheorem{lemma}{Lemma}

\newtheorem{remark}{Remark}
\newtheorem{assumption}{Assumption}

\begin{document}

\title{On the dynamics of two photons interacting with a two-qubit coherent feedback network}

\author{Guofeng Zhang\thanks{Department of Applied Mathematics, The Hong Kong Polytechnic University, Hung Hom, Kowloon,  Hong Kong (Guofeng.Zhang@polyu.edu.hk).} 
\and  Yu Pan\thanks{Institute of Cyber-Systems and Control, Zhejiang University, Hangzhou, 310027, China (ypan@zju.edu.cn).}
}
\maketitle

\begin{abstract}
The purpose of this paper is to study the dynamics of a quantum coherent feedback network composed of two two-level systems (qubits) driven by two counter-propagating photons, one in each input channel. The coherent feedback network enhances the nonlinear photon-photon interaction inside the feedback loop. By means of quantum stochastic calculus and the input-output framework, the analytic form of the steady-state output two-photon state is derived. Based on the analytic form, the applications on the Hong-Ou-Mandel (HOM) interferometer and marginally stable single-photon devices using this coherent feedback structure have been demonstrated. The difference between continuous-mode and single-mode few-photon states is demonstrated.
\end{abstract}

\textbf{keywords.} coherent feedback network; two-level systems; continuous-mode photon states

\section{Introduction}\label{sec:intro}

The last few decades have witnessed rapid advances in experimental demonstration and theoretical investigation of quantum control systems due to their promising applications in a wide range of areas such as quantum communication, quantum computing, and quantum metrology \cite{AA03}, \cite{MvH07}, \cite{JNP08}, \cite{BCS09}, \cite{LK09},  \cite{WM10}, \cite{DP10}, \cite{WS10}, \cite{AT12}, \cite{BQ13} \cite{ZJ13}, \cite{vHSM05}, \cite{NY17}, \cite{ZLW+17}. From a signals and systems point of view, quantum linear systems, prepared in Gaussian states and driven by Gaussian input states,  have been well studied; results like quantum filtering and measurement-based feedback control have been well established \cite{JNP08}, \cite{NJD09}, \cite{WM10}, \cite{NY17}. In addition to Gaussian states there are other types of non-classical states, for example single- and multi-photon states. Roughly speaking, a light field is in an $\ell$-photon state if there is a
definite number of $\ell$ photons in this field. A \textit{continuous-mode} $\ell$-photon state is characterized by
the frequency (or equivalently, temporal) profiles of these photons. Interaction between photons and quantum
finite-level systems has received considerable attention recently, as the
precise control of the interactions between photons and matter is fundamentally important for quantum information processing  \cite{kimble08}, \cite{Kolchin11}, \cite{LMS+17}, \cite{RWF17}. Two-photon interaction induced by finite-level systems is of particular interest since it introduces nonlinearity to the steady-state response.

\begin{figure}[tbp]
\centering
\includegraphics[scale=0.4]{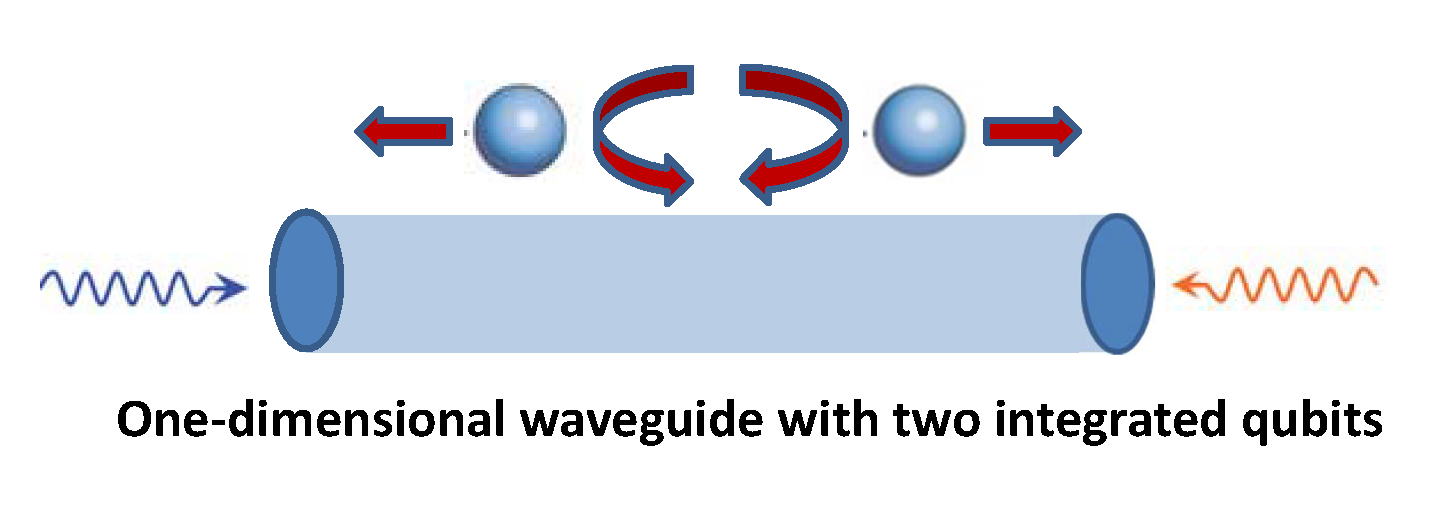}
\caption{Traveling photons are confined to a one-dimensional waveguide, which means they can only travel in two opposite directions. Due to the coupling between the photon and each qubit at the interaction location, the photon will either keep the original traveling direction or be reflected with probabilities that sum to 1, leading to a coherent feedback mechanism.}
\label{fig_cf}
\end{figure}

Photons do not interact in free space. Physically, the interaction can be mediated by quantum finite-level systems (quantum emitters). A simple example is the interaction of two photons by coupling to a qubit. In \cite{Shen07,Fan10}, two-photon transport properties have been studied by considering a one-dimensional waveguide coupled to a qubit. Intuitively, the response of the systems can be engineered by exploiting various configurations of quantum emitters. A scattering matrix analysis has shown that any localized quantum emitter inevitably induces frequency mixing and entanglement between two photons \cite{Fan10,SES13}. The response of  a two-level system to two continuous-mode photons has recently been investigated in \cite{DZA19}. These two photon can either co-propagate or counter-propagate along the waveguide. By means of a transfer function approach, the output field states for both cases are derived analytically.  The problem of two photons scattering off a two-level emitter residing in a 1D waveguide is studied in \cite{Nysteen:14}, where it is demonstrated that photon transport properties depend on the excitation of the emitter. Moreover,  correlations and entanglement between the two output  photons induced by a two-level emitter  are also investigated. The effect of the pulse shapes of two counter-propagating wavepackets on the correlations of the  output photons is studied in \cite{Nysteen15}, where the output intensity spectra are studied when the input photons are of Gaussian pulse shapes. In \cite{Roulet16}, both time and frequency correlations between output photons are discussed. Furthermore, the relationship between induced photon-photon correlations and the atomic excitation efficiency is investigated. When a two-level system is driven by two counter-propagating indistinguishable single photons, it is shown in \cite{DZA19} that the maximal excitation probability attains at $\gamma=5\kappa$ for rising exponential pulse shapes, and $\Omega=2*1.46\kappa$ for Gaussian pulse shapes. In \cite{SZX16}, quantum filters for a Markovian quantum system driven by an arbitrary number of photons in a single channel have been derived. Quantum filters are constructed in \cite{DZA19b} for a two-level atom driven by two counter-propagating photons, where numerical analysis reveals interesting scaling relations between atom-photon coupling strength and photonic pulse shape for maximum atomic excitation. The scattering of two photons  from two distant qubits embedded in a 1D waveguide has been studied in \cite{ZB13,Laasko14}. Due to the distance between the two emitters, non-Markovian effects exist in this setup, as numerically demonstrated by means of the Lippmannn-Schwinger equation in \cite{ZB13} and analytically investigated in terms of a Green function in \cite{Laasko14}. The generalizations to an array of multiple emitters can be found in \cite{YH15}. An experiment  that demonstrates photon-mediated qubit-qubit  interactions is performed in \cite{vLFL+13}.  By increasing the number of emitters, stronger photon-photon correlation can be obtained, which often provides more control options for generating entangled quantum states, engineering transmission properties, and synthesizing quantum gates \cite{ESS11,K16,BrodA16,Brod16,WJ17,SVF18}.  The reason for the stronger correlation with multi-qubit setting is that photons could interact multiple times at the emitters, or interact at several different sites \cite{BrodA16,Brod16}. As an application, a controlled-PHASE (CPHASE) quantum gate is proposed in \cite{Brod16}. Moreover, persistent oscillations of quantum correlations \cite{ZB13} have been observed if the two photons are allowed to bounce back after interacting with a qubit. Inspired by \cite{ZB13,Laasko14}, we investigate the steady-state response of a two-qubit system driven by two continuous-mode photons; Fig. \ref{fig_cf}. We model the system such that photons can be fed back after interacting with the other qubit, which provides a way for the photons to interact multiple times using a minimum number of qubits. This coherent feedback configuration could be realized using standard waveguide quantum electro-dynamics (QED) devices \cite{RWF17}, \cite{LMS+17}. One-dimensional waveguides can be realized in photonic nanostructures, or transmission lines in superconducting microwave circuits. Each qubit can be realized as an artificial superconducting circuit that is directly integrated with the waveguide. Alternatively, the qubit can be realized as an atomic ensemble or a single atom embedded in a cavity that is strongly coupled to the waveguide.

Numerical and analytical results have been obtained for a similar configuration which includes a feedback mechanism \cite{ZB13,Laasko14,YH15}. These previous works have considered two photons interacting with two distant qubits, which results in numerical and exact solutions characterizing spatial propagation of the photon wave functions. Non-Markovianity has also been considered in these works. In this paper, an alternative quantum network formalism is adopted \cite{HP84,GC85,GJ09,GZ00}. Based on Markovian quantum stochastic differential equations (QSDEs), this control-theoretic approach studies the steady-state response which captures the time-correlation of the output photons. Moreover, this formalism facilitates a network analysis which is applicable to any generic configuration. For example, the physical configuration in Fig. \ref{fig_cf} can be translated using the SLH language \cite{GJ09,CKS17} into a standard coherent feedback network structure; Fig. \ref{fig_sys}. The feedback network has two input channels, each containing one photon described in terms of its continuous-mode pulse shape. Two-photon scattering via a single qubit has been studied with the quantum network formalism \cite{Pan16,DZA19} or an equivalent input-output formalism \cite{Fan10,TMT15} before. However, the steady-state response has not been solved for a marginally stable system, which is our case. (The notion of ``a marginally stable system'' is interpreted in Remark \ref{rem:impulse function}.) In this paper, the steady-state output of a coherent feedback network with two continuous-mode photons as the input has been derived for the first time. A novel two-photon process has been found in the nonlinear response of the system. Based on the analytical results of the system response, it is possible to synthesize photonic systems to achieve desirable dynamics using the enhanced nonlinearity. We have demonstrated the results with a tunable HOM design and a marginally-stable single-photon device.

The rest of the paper is organized as follows.  The coherent quantum feedback network and two-photon input state are introduced in Sec. \ref{sec:pre}.  The main result of this paper, an analytic form of the steady-state output two-photon state,  is presented in Sec. \ref{sec:main_result}. Two direct applications are discussed in Sec. \ref{sec:app}. Sec. \ref{sec:conclusion} concludes this paper.

{\it Notation.}    $x^{\ast }$ denotes the complex conjugate of a complex number $x$ or
the adjoint of an operator $x$. The commutator of two operators $X$ and $Y$
is defined as $[X,Y]\triangleq XY-YX$. For a column vector $X=[x_{i}]$ with number or operator entries, $
X^{\#}=[x_{i}^{\ast }]$  and $X^\dag=(X^\#)^T$. $I_{k}$ is the identity matrix and $0_{k}$ the zero matrix in $
\mathbb{C}^{k\times k}$. $\delta _{ij}$ denotes the Kronecker delta and $\delta(t)$ denotes the Dirac delta.

\section{Coherent feedback network and input state} \label{sec:pre}

\begin{figure}[tbp]
\centering
\includegraphics[scale=0.8]{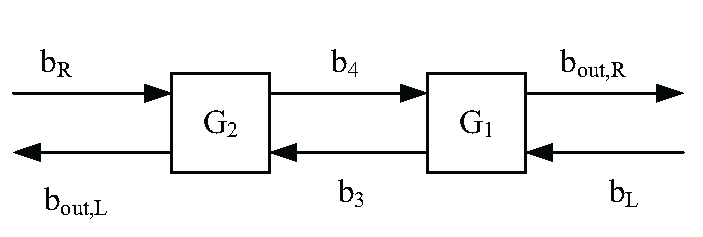}
\caption{$G_1$ and $G_2$ are two-level systems. The coherent feedback network is driven by two photons, one in each input channel designated by $b_{\rm L}$ and $b_{\rm R}$ respectively. $b_{\rm out,L}$ and $b_{\rm out,R}$ denote the two output channels.}
\label{fig_sys}
\end{figure}

In this section, we introduce the coherent feedback network, as is shown in Fig. \ref{fig_sys}. We also introduce the two-photon input state for this feedback network.

\subsection{Coherent feedback network}\label{subsec:network}
The open quantum system under study can be properly modeled using a triplet $(S,L,H)$ \cite{HP84,GJ09,ZJ12,CKS17}. Here, $S$ is a scattering operator, and the system is coupled to the photonic fields through the operator $L$. $H$ is the inherent Hamiltonian of the system. The overall dynamics of an open quantum system interacting with the input fields is governed by a unitary operator $U(t,t_0)$, where $t_0$ is the initial time of the interaction. The dynamical equation of $U(t,t_0)$, $t\geq t_0$, is given by \cite{HP84}
\begin{equation}\label{lem1p}
dU(t,t_0)=\{b^\dag(t)L-L^\dag Sb(t)-(\frac{1}{2}L^\dag L+iH)\}U(t,t_0)dt
\end{equation}
with $U(t+dt,t_0) = U(t,t_0) + dU(t,t_0)$ and $U(t_0,t_0)=I\otimes I$ being the identity operator of the composite system. $b(t)$ is a vector of annihilation operators
for the input field modes. Physically, $b(t)$ and $b^\#(t)$ can be understood as the annihilation and
creation of photons in the fields at time $t$. Note that Markovian approximation has been invoked in the derivation of $dU(t,t_0)$. The Heisenberg-picture evolution of a system operator $X$ can be calculated by $X(t)=U^*(t,t_0)(X\otimes I)U(t,t_0)$, with $I$ being the identity operator on the fields. The dynamical equation of $X(t)$ is then given by the following QSDE \cite{HP84,GJ09}
\begin{eqnarray}
\dot{X}(t)&=&\mathcal L^*(X(t))
\nonumber
\\
&&+b^\dag(t)S^\dag[X(t),L(t)]+[L^\dag(t),X(t)]Sb(t)
\label{eq:X Feb24}
\end{eqnarray}
where
\begin{eqnarray*}
\mathcal L^*(X(t))&\triangleq&
-i[X(t),H(t)]+L^\dag(t) X(t)L(t)\\
&&
-\frac{1}{2}L(t)^\dag L(t)X(t)-\frac{1}{2}X(t)L^\dag(t) L(t)
\end{eqnarray*}
iis the Lindblad operator and the other two are noise terms.  Moreover, the output $b_{\rm out}(t)$ is related to the input $b(t)$ via the following relation \cite{GHN+12}
\begin{equation}\label{inout}
b_{\rm out}(t)=U^*(t,t_0)(I\otimes b(t))U(t,t_0),
\end{equation}
whose dynamics is given by
\[
b_{\rm out}(t)=L(t)+Sb(t).
\]
The coherent feedback network, as shown in Fig. \ref{fig_sys},  has two inputs represented by annihilation operators $b_{\rm L}$ and $b_{\rm R}$ respectively.  $G_1$ and $G_2$ are two-level systems, whose ground and excited state vectors   are  $\ket{g_j}$ and $\ket{e_j}$ ($j=1,2$) respectively.
\begin{assumption}\label{assume1}
The coherent feedback network in Fig. \ref{fig_sys} is assumed to satisfy the following conditions.
\begin{itemize}
\item The central frequencies of the two input fields $b_{\rm L}$ and $b_{\rm R}$ are the same, denoted by $\omega _{o}$.
\item  $G_1$ and $G_2$ have the same transition frequency between the ground state and excited state, denoted by \  $\omega_{a}$. Thus, the detuning frequency is $\omega _{c}=\omega _{o}-\omega _{a}$.
\item $G_1$ and $G_2$ have the same coupling strength $\kappa$ to the optical fields.
\end{itemize}
\end{assumption}
Under {\bf Assumption \ref{assume1}}, the triplet $(S,L,H)$ for the two-level systems are given by \cite{GJ09,ZJ12,CKS17,NY17}
\begin{equation}\label{sys0}
G_{j}=\left(I_2,\left[
\begin{array}{c}
1 \\
1
\end{array}
\right] \sqrt{\kappa }\sigma _{-,j},\frac{\omega _{c}}{2}\sigma_{z,j}\right) ,\ j=1,2, 
\end{equation}
where $\sigma_{-,j}=\ket{g_j}\bra{e_j}$ is the lowering ladder operator and $\sigma_{z,j}=\ket{e_j}\bra{e_j}-\ket{g_j}\bra{g_j}$ is the Pauli $Z$ operator for the system $G_j$. As the coupling operators are the lowering ladder operators,  the two-level systems $G_1$ and $G_2$  undergo amplitude damping, \cite[Chapter 8]{NC10}.

As can be seen from the above equation, there are two coupling channels for each $G_j$, which model the interaction with the left-going and right-going photons.
Recall that $t_0$ is  the time when the system and its inputs start to interact. What we are interested in this paper is the steady-state dynamics of the coherent feedback network in the limit $t_{0}\rightarrow -\infty $ and $t\to \infty$; i.e.,  the interaction occurs in the remote past and we look at the dynamics in the distant future; see e.g., \cite{Fan10}, \cite{ZJ13}, \cite{YJ14},  \cite{Pan16}. Define
\begin{equation}
\alpha \triangleq -i\omega _{c}-\kappa .  \label{alpha}
\end{equation}
Substituting  the $(S,L,H)$ parameters in \eqref{sys0} into the system equation \eqref{eq:X Feb24} and the input-output relation \eqref{inout}, it can be readily shown that the QSDEs for the two-level system $G_{1}$ are
\begin{eqnarray*}
\dot{\sigma}_{-,1}(t)& =&\alpha \sigma _{-,1}(t)+\sqrt{\kappa }\sigma
_{z,1}(t)b_{\rm L}(t)+\sqrt{\kappa }\sigma _{z,1}(t)b_{4}(t),  \\
b_{3}(t)& =&\sqrt{\kappa }\sigma _{-,1}(t)+b_{\rm L}(t),  \\
b_{\mathrm{out,R}}(t)& =&\sqrt{\kappa }\sigma _{-,1}(t)+b_{4}(t),\ \ \ t\geq
t_{0}.
\end{eqnarray*}
Likewise, the QSDEs for the two-level system $G_{2}$ are
\begin{eqnarray*}
\dot{\sigma}_{-,2}(t)& =&\alpha \sigma _{-,2}(t)+\sqrt{\kappa }\sigma
_{z,2}(t)b_{3}(t)+\sqrt{\kappa }\sigma _{z,2}(t)b_{\rm R}(t),
\\
b_{\mathrm{out,L}}(t)& =&\sqrt{\kappa }\sigma _{-,2}(t)+b_{3}(t),
 \\
b_{4}(t)& =&\sqrt{\kappa }\sigma _{-,2}(t)+b_{\rm R}(t),\ \ \ t\geq t_{0}.
\end{eqnarray*}
Consequently, the QSDEs for the coherent feedback network in  Fig. \ref{fig_sys}  are
\begin{subequations}
\begin{align}
\left[
\begin{array}{c}
\dot{\sigma}_{-,1}(t) \\
\dot{\sigma}_{-,2}(t)
\end{array}
\right] & =\alpha \left[
\begin{array}{c}
\sigma _{-,1}(t) \\
\sigma _{-,2}(t)
\end{array}
\right] +\kappa \left[
\begin{array}{c}
\sigma _{z,1}(t)\sigma _{-,2}(t) \\
\sigma _{z,2}(t)\sigma _{-,1}(t)
\end{array}
\right]  \nonumber \\
&\ \ \  +\sqrt{\kappa }\left[
\begin{array}{c}
\sigma _{z,1}(t) \\
\sigma _{z,2}(t)
\end{array}
\right] \left( b_{\rm L}(t)+b_{\rm R}(t)\right) ,  \label{sys_e} \\
\nonumber \\
b_{\mathrm{out}}(t)& =\sqrt{\kappa }\ C \left[
\begin{array}{c}
\sigma _{-,1}(t) \\
\sigma _{-,2}(t)
\end{array}
\right] +b_{\mathrm{in}}(t),  \  t\geq t_{0},  \label{sys_f}
\end{align}
\end{subequations}
where
\begin{equation}
C = \left[
\begin{array}{cc}
1 & 1 \\
1 & 1
\end{array}
\right] ,  \label{C}
\end{equation}
 and
\begin{equation*}
b_{\mathrm{in}}(t)\triangleq \left[
\begin{array}{c}
b_{\rm L}(t) \\
b_{\rm R}(t)
\end{array}
\right] ,\ b_{\mathrm{out}}(t)\triangleq \left[
\begin{array}{c}
b_{\mathrm{out,L}}(t) \\
b_{\mathrm{out,R}}(t)
\end{array}
\right]
\end{equation*}
are input and output fields for the feedback network respectively.

In what follows, we present the Fourier transform of operators and functions to be used in the sequel. For the vector of inputs $b_{\mathrm{in}}(t)$ in the time domain, we define its Fourier
transform as
\begin{equation}
b_{\mathrm{in}}[i\omega ]
\triangleq
\frac{1}{\sqrt{2\pi }}\int_{t_{0}}^{\infty }dt\ e^{-i\omega t}b_{\mathrm{in}}(t),
\ \ \ \omega \in \mathbb{R}.
\label{b_s_1}
\end{equation}
The inverse Fourier Transform is
\begin{equation}
b_{\mathrm{in}}(t) = \frac{1}{\sqrt{2\pi }}\int_{-\infty}^{\infty }d\omega\ e^{i\omega t}b_{\mathrm{in}}[i\omega],\ \ \ \  t\geq t_0.
\label{b_t_1}
\end{equation}

\begin{remark}\label{rem:t_0}
As mentioned above,  the initial time $t_0$ will be sent to $-\infty$ later, thus Eq. (\ref{b_s_1}) is indeed the conventional Fourier transform. The same is true for  the Fourier transform of other operators or functions to be presented in the sequel.
\end{remark}

 The adjoint $b_{\mathrm{in}}^{\dag}[i\omega ]$ of $b_{\mathrm{in}}[i\omega] $ is obtained by conjugating both sides of Eq. (\ref{b_s_1}), specifically,
\begin{equation}
b_{\mathrm{in}}^{\dag}[i\omega ]
=
\frac{1}{\sqrt{2\pi }}\int_{t_{0}}^{\infty }dt\ e^{i\omega t}b_{\mathrm{in}}^{\dag}(t),
\ \ \ \omega \in \mathbb{R}.
\label{b_s_2}
\end{equation}
Noticing
\begin{equation}
\lim_{t_{0}\rightarrow -\infty }\frac{1}{2\pi }\int_{t_{0}}^{\infty }dt\
e^{i\omega t} =\delta (\omega ),
\label{july2_delta}
\end{equation}
and the commutation relation
\begin{equation}
\lbrack b_{\mathrm{in}}(t),b_{\mathrm{in}}^{\dag}(r)]=\delta (t-r)I_2,\ \ \
t,r\geq t_{0},  \label{CCR_t}
\end{equation}
we have that for arbitrary $\omega _{1},\omega _{2}\in \mathbb{R}$,
\begin{align*}
&
 \lim_{t_{0}\rightarrow -\infty }[b_{\mathrm{in}}[i\omega _{1}],\ b_{\mathrm{in}}^{\dag}[i\omega _{2}]]
\\
=&
\lim_{t_{0}\rightarrow -\infty }\frac{1}{2\pi }\int_{t_{0}}^{\infty }dt\  e^{-i(\omega _{1}-\omega _{2})t} I_2
=
\delta (\omega _{1}-\omega _{2}) I_2 .
 \label{CCR_f}
\end{align*}
Similarly, we denote the Fourier transform of the vector of outputs $b_{\rm out}(t)$ by $b_{\rm out}[i\omega]$, whose  adjoint is denoted  by $b_{\rm out}^\dag [i\omega]$. Finally, the Fourier transform of $\sigma _{-}(t)$ is
\begin{equation}
\sigma _{-}[i\omega ]
=
 \frac{1}{\sqrt{2\pi }}\int_{t_{0}}^{\infty }dt\ e^{-i\omega t}\sigma _{-}(t),
\label{july11_1a}
\end{equation}
whose adjoint is
\begin{equation}
\sigma _{+}[i\omega ]
=
(\sigma _{-}[i\omega ])^{\ast }=\frac{1}{\sqrt{2\pi }}\int_{t_{0}}^{\infty }dt\ e^{i\omega t}\sigma _{+}(t),
\label{july11_1b}
\end{equation}
where $\sigma _{+}(t)$,  the raising ladder operator, is the adjoint of $\sigma _{-}(t)$.

\subsection{Two-photon input state}\label{subsec:state}
In this subsection, we introduce the input to the feedback network in Fig. \ref{fig_sys}. The left-going input field is in the continuous-mode single-photon state $b_{\rm L}^{\ast }(\xi_{\rm L})\ket{0_{\rm L}}$, where $\ket{0_{\rm L}}$ denotes the vacuum state of this field, and the operator  $b_{\rm L}(\xi_{\rm L}) $ is defined to be
\[
b_{\rm L}(\xi_{\rm L}) \triangleq \int_{t_{0}}^{\infty }b_{\rm L}(t)\xi_{\rm L}
^{\ast }(t)dt
\]
with  $\xi_{\rm L}\in L_{2}(\mathbb{R},\mathbb{C})$ satisfying the normalization condition   $\left\Vert \xi_{\rm L}\right\Vert \equiv \sqrt{\int_{t_0}^\infty |\xi_{\rm L}(t)|^2 dt}=1$. In other words, $\|\xi_{\rm L}\|=1$ guarantees that the state $b_{\rm L}^{\ast }(\xi_{\rm L})\ket{0_{\rm L}}$ is normalized. The physical interpretation of $\xi(t)$ is that $|\xi(t)|^2 dt$ is the probability of finding the photon in the time interval $[t, t+dt)$.  Similarly, the right-going input field is in the continuous-mode single-photon state $b_{\rm R}^{\ast }(\xi_{\rm R})\ket{0_{\rm R}}$, where $\ket{0_{\rm R}}$ denotes the vacuum state of this field, and the operator  $b_{\rm R}(\xi_{\rm R} ) $ is defined to be
\[
b_{\rm R}(\xi_{\rm R} ) \triangleq \int_{t_{0}}^{\infty }b_{\rm R}(t)\xi_{\rm R}
^{\ast }(t)dt
\]
with $\xi_{\rm R} \in L_{2}(\mathbb{R},\mathbb{C})$ being the temporal pulse function of the photon and satisfying $\left\Vert \xi_{\rm R}\right\Vert=1$. The adjoints of $b_{\rm L}(\xi_{\rm L})$ and  $b_{\rm R}(\xi_{\rm R})$  are
\begin{subequations}
\begin{eqnarray}
b_{\rm L}^{\ast }(\xi_{\rm L})
&\triangleq&
 (b_{\rm L}(\xi_{\rm L}))^\ast =   \int_{t_{0}}^{\infty }b_{\rm L}^{\ast}(t)\xi_{\rm L}(t)dt,
 \label{may22_B_LR0}
  \\
b_{\rm R}^{\ast }(\xi_{\rm R} )
&\triangleq&
(b_{\rm R}(\xi_{\rm R} ))^\ast = \int_{t_{0}}^{\infty }b_{\rm R}^{\ast}(t)\xi_{\rm R} (t)dt,
 \label{may22_B_LR}
\end{eqnarray}
\end{subequations}
respectively. Thus, the two-photon input field state is
\begin{equation}
\ket{\Psi _{\mathrm{in}}(t_{0})}
=
b_{\rm L}^{\ast }(\xi_{\rm L})b_{\rm R}^{\ast }(\xi_{\rm R})\ket{0_{\rm L}0_{\rm R}} .
\label{initial}
\end{equation}
Similar to Eq. (\ref{b_s_1}), the Fourier transform of a function $\xi \in L_{2}(\mathbb{R},\mathbb{C})$ is
\begin{equation}\label{xi_nu}
\xi[i\nu ]=\frac{1}{\sqrt{2\pi }}\int_{t_{0}}^{\infty }dt\ e^{-i\nu t}\xi(t),
\end{equation}
whose inverse Fourier transform is
\begin{equation}\label{xi_t}
\xi(t)=\frac{1}{\sqrt{2\pi }}\int_{-\infty}^{\infty }d\nu\ e^{i\nu t}\xi[i\nu],
\  \ \ t\geq t_0.
\end{equation}

\begin{example}\label{ex:photon}
 For the purpose of demonstration, we consider two single-photon states of Lorentzian-type pulse shape
\begin{equation}
\xi _{j}[i\nu ]=\frac{1}{\sqrt{2\pi }}\frac{\sqrt{\gamma _{j}}}{i(\nu
-\omega _{o})-\frac{\gamma _{j}}{2}},\ \ j= {\rm L,R},  \label{dec9_xi_f}
\end{equation}
which in the time-domain are
\begin{equation}
\xi _{j}(t)=\left\{
\begin{array}{cc}
0, & t\geq 0, \\
-\sqrt{\gamma _{j}}e^{(\frac{\gamma _{j}}{2}+i\omega _{o})t}, & t<0,
\end{array}
\right. ,\ \ j= {\rm L,R}.  \label{dec9_xi_t}
\end{equation}
Here, $\omega_o$ is the central frequency of the fields, as discussed in {\bf Assumption 1}.  In particular,  when $\gamma _{\rm L}=\gamma _{\rm R} =\gamma$,  the two photons have the same pulse shape $\xi_{\rm L} = \xi_{\rm R} \equiv \xi
$, given by
\begin{equation}\label{jan14_xi}
\xi \lbrack i\nu ]=\frac{1}{\sqrt{2\pi }}\frac{\sqrt{\gamma }}{i(\nu-\omega_o) -\frac{\gamma }{2}}.
\end{equation}
For Lorentzian-type pulse shapes, $\gamma $ is commonly called the full width at half maximum (FWHM); see, e.g., \cite[Chapter 2]{RL00}. It has been shown that a  Lorentzian-type single photon, which has a temporal  pulse shape of the form \eqref{dec9_xi_t}, is able to excite a two-level atom fully;  see, e.g.,  \cite{WMS+11}, \cite{YJ14}, \cite{PZJ15}.
\end{example}

More discussions on continuous-mode single- and multi- photon states can be found in, e.g., \cite{RL00}, \cite{LMS+17}, \cite{Z19}.

\section{Steady-state output field state} \label{sec:main_result}

In this section, we derive the steady-state output field state of the 2-qubit coherent feedback network driven by two photons, as described in the previous section.

\subsection{Basic set-up} \label{subsec:setup}

Let the two-level systems $G_1$ and $G_2$ be initialized in the ground states $|g_1\rangle$ and $|g_2\rangle$ respectively, and the input be in the two-photon state as given in Eq. (\ref{initial}). The abbreviation $|0\rangle=|g_1g_2\rangle$ is used for the ground state.

\begin{assumption}\label{assume2}
The initial joint system-field state is
\begin{equation*}
\ket{\Psi (t_{0})}
=
\ket{\Psi _{\mathrm{in}}(t_{0})}\ket{0}
=
b_{\rm L}^{\ast }(\xi_{\rm L})b_{\rm R}^{\ast }(\xi_{\rm R})\ket{0_{\rm L}0_{\rm R}0}.
\label{Psi_t0}
\end{equation*}
\end{assumption}
In  the Schr\"{o}dinger picture, the system-field state undergoes a unitary evolution. At time instant $t\geq t_{0}$, the joint system-field state is
\begin{equation*}
\ket{\Psi (t)}
=
U(t, t_{0})\ket{\Psi (t_{0})}.
\label{t}
\end{equation*}
In the steady-state limit ($t_{0}\rightarrow -\infty ,t\rightarrow \infty $), the photons are in the two output channels, leaving the two-level
systems in their ground state. Then the steady-state output field state $\ket{\Psi _{\mathrm{out}}}$ can be obtained by tracing out the system state; i.e.,
\begin{equation}
\ket{\Psi _{\mathrm{out}}}
=
\lim_{t_{0}\rightarrow -\infty ,t\rightarrow \infty }\braket{
0|\Psi (t)}. \label{infty}
\end{equation}
As the system-field interaction does not generate photons, i.e. the combined system is passive, $\ket{\Psi _{\mathrm{out}}}$ is a two-photon state with the time-domain basis
\begin{align}
&\bigg\{\frac{1}{2}\int_{-\infty}^{\infty }dp_{1}\int_{-\infty}^{\infty }dp_{2}\
\ket{1_{{\rm L}p_{1}}1_{{\rm L}p_{2}}}
\bra{1_{{\rm L}p_{1}}1_{{\rm L}p_{2}}} ,
\nonumber
 \\
&\int_{-\infty}^{\infty }dp_{1}\int_{-\infty}^{\infty }dp_{2}\ \ket{1_{{\rm L}p_{1}}1_{{\rm R}p_{2}}} \bra{1_{{\rm L}p_{1}}1_{{\rm R}p_{2}}},  \nonumber \\
&\ \frac{1}{2}\int_{-\infty}^{\infty }dp_{1}\int_{-\infty}^{\infty }dp_{2}\
\ket{1_{{\rm R}p_{1}}1_{{\rm R}p_{2}}} \bra{1_{{\rm R}p_{1}}1_{{\rm R}p_{2}}}\bigg\},
\label{dec6:basis}
\end{align}
where the notation
\begin{equation}
\ket{1_{jt}} \equiv b_j^{\ast }(t)\ket{0_j},\ \ t\geq t_{0}, \ \ j = {\rm L, R}
  \label{impulse}
\end{equation}
denotes the instantaneous generation of a left-going (right-going) photon at time $t$ in the fields. By inserting Eq. (\ref{dec6:basis}) into the RHS of Eq. (\ref{infty}) and noticing Eq. (\ref{inout}), we obtain
\begin{align}
&\ket{\Psi _{\mathrm{out}}}
\label{mar17_1}
 \\
=&
 \lim_{t_0\to-\infty}\frac{1}{2}\int_{-\infty}^\infty dp_{1}\int_{-\infty}^\infty dp_{2}   \ket{1_{{\rm L}p_{1}}1_{{\rm L}p_{2}}}
 \nonumber\\
 &\times\int_{t_{0}}^{\infty}dt_{1}\int_{t_{0}}^{\infty }dt_{2}\ \xi_{\rm L}(t_{1})\xi_{\rm R}(t_{2})&
 \nonumber
\\
&
 \times \left\langle 0_{\rm L}0_{\rm R}0\right\vert b_{\mathrm{out,L}}(p_{1})b_{\mathrm{out,L}}(p_{2})b_{\rm L}^{\ast }(t_{1})b_{\rm R}^{\ast}(t_{2})\left\vert 0_{\rm L}0_{\rm R}0\right\rangle
\nonumber
 \\
&
+ \lim_{t_0\to-\infty}\int_{-\infty}^\infty dp_{1}\int_{-\infty}^\infty dp_{2}   \ket{1_{{\rm L}p_{1}}1_{{\rm R}p_{2}}}
\nonumber
\\
& \times\int_{t_{0}}^{\infty }dt_{1}\int_{t_{0}}^{\infty }dt_{2}\ \xi_{\rm L}(t_{1})\xi
_{\rm R}(t_{2})
\nonumber
\\
&
 \times \left\langle 0_{\rm L}0_{\rm R}0\right\vert b_{\mathrm{out
,L}}(p_{1})b_{\mathrm{out,R}}(p_{2})b_{\rm L}^{\ast }(t_{1})b_{\rm R}^{\ast
}(t_{2})\left\vert 0_{\rm L}0_{\rm R}0\right\rangle
\nonumber
 \\
&
+ \lim_{t_0\to-\infty}\frac{1}{2}\int_{-\infty}^\infty dp_{1}\int_{-\infty}^\infty dp_{2} \ket{1_{{\rm R}p_{1}}1_{{\rm R}p_{2}}}
\nonumber
\\
&\times \int_{t_{0}}^{\infty}dt_{1}\int_{t_{0}}^{\infty }dt_{2} \  \xi_{\rm L}(t_{1})\xi_{\rm R}(t_{2})
\nonumber
\\
&
\times \langle 0_{\rm L}0_{\rm R}0\vert b_{\mathrm{out,R}}(p_{1})b_{\mathrm{out,R}}(p_{2})b_{\rm L}^{\ast }(t_{1})b_{\rm R}^{\ast}(t_{2})\vert 0_{\rm L}0_{\rm R}0\rangle. \nonumber
\end{align}
Note that since $U(t,t_0)|0_{\rm L}0_{\rm R}0\rangle=\theta(t)|0_{\rm L}0_{\rm R}0\rangle$, $|\theta(t)|=1$ holds by Lemma 3 in \cite{PZJ15}, an irrelevant overall phase factor $\theta(t)$ has been omitted in Eq. (\ref{mar17_1}). Next, we go to the frequency domain by applying the Fourier
transform to the time variables $t_{1},t_{2}$ and $p_1,p_2$, respectively. According to Eqs. (\ref{b_t_1}) and (\ref{xi_t}), we have the following frequency-domain counterpart of Eq. \eqref{mar17_1}
\begin{align}
&\ket{\Psi _{\mathrm{out}}}
 \label{dec8_Psi}
  \\
=& \frac{1}{2}\int\limits_{\omega_1,\omega_2,\nu_1,\nu_2}\xi_{\rm L}[i\nu _{1}]\xi_{\rm R}[i\nu
_{2}]\ket{1_{{\rm L}\omega_{1}}1_{{\rm L}\omega _{2}}}
 \nonumber
 \\
& \ \ \ \times
\left\langle 0_{\rm L}0_{\rm R}0\right\vert b_{\mathrm{out,L}}[i\omega
_{1}]b_{\mathrm{out,L}}[i\omega _{2}]b_{\rm L}^{\ast }[i\nu _{1}]b_{\rm R}^{\ast
}[i\nu _{2}]\left\vert 0_{\rm L}0_{\rm R}0\right\rangle
 \nonumber
  \\
& +\int\limits_{\omega_1,\omega_2,\nu_1,\nu_2}\xi_{\rm L}[i\nu _{1}]\xi_{\rm R}[i\nu
_{2}]\ket{1_{{\rm L}\omega_{1}}1_{{\rm R}\omega _{2}}}
 \nonumber
  \\
&\ \ \ \times \left\langle 0_{\rm L}0_{\rm R}0\right\vert b_{\mathrm{out,L}}[i\omega
_{1}]b_{\mathrm{out,R}}[i\omega _{2}]b_{\rm L}^{\ast }[i\nu _{1}]b_{\rm R}^{\ast
}[i\nu _{2}]\left\vert 0_{\rm L}0_{\rm R}0\right\rangle
\nonumber
 \\
& +\frac{1}{2}\int\limits_{\omega_1,\omega_2,\nu_1,\nu_2}\xi_{\rm L}[i\nu _{1}]\xi_{\rm R}[i\nu
_{2}]\ket{1_{{\rm R}\omega_{1}}1_{{\rm R}\omega _{2}}}
 \nonumber
  \\
& \ \ \ \times   \left\langle 0_{\rm L}0_{\rm R}0\right\vert b_{\mathrm{out,R}}[i\omega
_{1}]b_{\mathrm{out,R}}[i\omega _{2}]b_{\rm L}^{\ast }[i\nu _{1}]b_{\rm R}^{\ast
}[i\nu _{2}]\left\vert 0_{\rm L}0_{\rm R}0\right\rangle ,  \nonumber
\end{align}
where we have used the abbreviation
\begin{equation*}
\int\limits_{\omega_1,\omega_2,\nu_1,\nu_2}\equiv\int_{-\infty}^\infty d\omega _{1}\int_{-\infty}^\infty d\omega _{2}\int_{-\infty }^{\infty }d\nu
_{1}\int_{-\infty }^{\infty }d\nu _{2}.
\end{equation*}
Hence, in order to find an analytical expression for $\ket{\Psi _{\mathrm{out}}}$, we have to calculate the following quantities:
\begin{subequations}
\begin{align}
& \langle 0_{\rm L}0_{\rm R}0\vert b_{\mathrm{out,L}}[i\omega
_{1}]b_{\mathrm{out,L}}[i\omega _{2}]b_{\rm L}^{\ast }[i\nu _{1}]b_{\rm R}^{\ast
}[i\nu _{2}]\vert 0_{\rm L}0_{\rm R}0\rangle,
\label{3_key_a}
\\
&\langle 0_{\rm L}0_{\rm R}0\vert b_{\mathrm{out,L}}[i\omega
_{1}]b_{\mathrm{out,R}}[i\omega _{2}]b_{\rm L}^{\ast }[i\nu _{1}]b_{\rm R}^{\ast
}[i\nu _{2}]\vert 0_{\rm L}0_{\rm R}0\rangle,
 \label{3_key_b}
\\
&\langle 0_{\rm L}0_{\rm R}0\vert b_{\mathrm{out,R}}[i\omega
_{1}]b_{\mathrm{out,R}}[i\omega _{2}]b_{\rm L}^{\ast }[i\nu _{1}]b_{\rm R}^{\ast
}[i\nu _{2}\vert 0_{\rm L}0_{\rm R}0\rangle.
\label{3_key_d}
\end{align}
\end{subequations}
The terms (\ref{3_key_a})-(\ref{3_key_d}) characterize the input-output relation of two photons. For example, Eq. (\ref{3_key_a}) characterizes the process of transferring two input counter-propagating photons at frequencies $\nu_1,\nu_2$ to two left-going output photons at frequencies $\omega_1,\omega_2$. The expression of the steady-state output field state $\ket{\Psi _{\mathrm{out}}}$ in Eq. \eqref{dec8_Psi} is fairly complicated. The purpose of the next subsection is to present a much simpler version of it; cf. Theorem \ref{thm:output_state}.

\subsection{The steady-state output state} \label{subsec:main}

Define a matrix
\begin{equation}
A\triangleq \left[
\begin{array}{cc}
\alpha & -\kappa \\
-\kappa & \alpha%
\end{array}
\right] =-\left[
\begin{array}{cc}
i\omega _{c}+\kappa & \kappa \\
\kappa & i\omega _{c}+\kappa
\end{array}
\right] ,  \label{A}
\end{equation}
where $\alpha$ is given in Eq. (\ref{alpha}).
It is easily found that the eigenvalues of the matrix $A$ are $-2\kappa
-i\omega _{c}$ and $-i\omega _{c}$.  Clearly, $A$ is marginally stable as it has an imaginary eigenvalue. $A$ being only marginally stable has a great impact on the derivation of Eq. (\ref{3_key_a})-(\ref{3_key_d}). To be more specific, the standard procedure to solve Eq. (\ref{3_key_a})-(\ref{3_key_d}) is to relate $b_\mathrm{out}(t)$ which is the time-domain counterpart of $b_\mathrm{out}[i\omega]$, to $b_\mathrm{in}(t)$ via Eq. (\ref{sys_f}), see, e.g., \cite{Fan10}. Then the remaining task is to solve the corresponding dynamics of $\sigma_{-,i}(t)$. For example, the following equation is a key part in the derivation
\begin{align}
&\bra{0_{\rm L}0_{\rm R}0} \left[
\begin{array}{c}
\dot{\sigma}_{-,1}(t) \\
\dot{\sigma}_{-,2}(t)
\end{array}
\right]
\label{sol}
 \\
=&A\bra{0_{\rm L}0_{\rm R}0} \left[
\begin{array}{c}
\sigma _{-,1}(t) \\
\sigma _{-,2}(t)
\end{array}
\right] -\sqrt{\kappa }C\bra{0_{\rm L}0_{\rm R}0} b_{
\mathrm{in}}(t),
\nonumber
\end{align}
which is based on Eq. (\ref{sys_e}) and the fact that
\begin{eqnarray}
&&\bra{0_{\rm L}0_{\rm R}0} \sigma _{z,i}(t)=\bra{0_{\rm L}0_{\rm R}0}U^\dagger(t,t_0)\sigma _{z,i}U(t,t_0)\nonumber\\
&=&-\bra{0_{\rm L}0_{\rm R}0}.  \label{july14_7}
\end{eqnarray}

Integrating both sides of Eq. (\ref{sol}) from $t_{0}$ to $t$ yields
\begin{eqnarray}
&&\bra{0_{\rm L}0_{\rm R}0} \left[
\begin{array}{c}
\sigma _{-,1}(t) \\
\sigma _{-,2}(t)
\end{array}
\right]
\nonumber
 \\
&=&e^{A(t-t_{0})}\bra{0_{\rm L}0_{\rm R}0} \left[
\begin{array}{c}
\sigma _{-,1}(t_{0}) \\
\sigma _{-,2}(t_{0})
\end{array}
\right]
 \nonumber
 \\
&-&\int_{t_{0}}^{t}d\tau\ \sqrt{\kappa }e^{A(t-\tau)}C\bra{
0_{\rm L}0_{\rm R}0} b_{\mathrm{in}}(\tau).
\label{sol_1}
\end{eqnarray}
When $A$ is Hurwitz, the first term in the RHS of Eq. (\ref{sol_1}) can be removed by taking the steady-state limit $t_0\rightarrow-\infty$. However, since $A$ is only marginally stable in our case, the initial time constant $t_0$ has to be included in the calculation and the steady-state limit can only be taken when appropriate. Furthermore, in contrast to the Hurwitz stable case, quantum It\^{o} calculus has to be explicitly invoked for the calculation of the nonlinear terms, which can be seen from the proof of Lemma \ref{lem:key} in the APPENDIX.

For later use, we define a matrix function
\begin{equation}
g_{G}(t)\triangleq \left\{
\begin{array}{cc}
\delta (t)-\kappa Ce^{At}C, & t\geq 0, \\
0, & t<0.
\end{array}
\right.
 \label{sept7_tf}
\end{equation}

\begin{remark}\label{rem:impulse function}
Formally, Equations (\ref{sol}) and (\ref{sys_f})  define a linear system with system matrices $(A, -\sqrt{\kappa}C, \sqrt{\kappa}C)$. In this sense,  $g_{G}(t)$ defined in Eq. (\ref{sept7_tf}) is of the form of an impulse response function, which is very commonly used in classical linear systems theory. Actually,   impulse response functions play an important role in quantum linear systems theory, see, e.g., \cite[Chapter 7]{WM08}, \cite{JNP08}, \cite[Chapter 6]{WM10}, \cite{ZJ13}, \cite{YJ14}, \cite{PZJ15}, \cite{NY17}, \cite{ZGPG18} and references therein. Moreover, since the matrix $A$  in Eq. (\ref{A}) is marginally stable, the linear system given by  equations (\ref{sol}) and (\ref{sys_f}) is marginally stable in the sense of linear systems theory. With slight abuse of notation,   we also say that our coherent feedback network in Fig. \ref{fig_sys} is marginally stable.
\end{remark}

For the time domain function $g_{G}(t)$ defined in Eq. (\ref{sept7_tf}), we
define its Laplace transform to be
\begin{equation}
G[s] \triangleq \int_{0}^{\infty }dt\ g_{G}(t)e^{-st}.  \label{L}
\end{equation}
Substituting Eq. \eqref{sept7_tf} into Eqs. (\ref{L}), we obtain
\begin{align}
&G[i\omega ]
=\frac{1}{\omega +\omega _{c}-2i\kappa }\left[
\begin{array}{cc}
\omega +\omega _{c} & 2i\kappa \\
2i\kappa & \omega +\omega _{c}
\end{array}
\right]
\nonumber
 \\
\triangleq &\left[
\begin{array}{c}
\Theta_{\rm L}[i\omega ] \\
\Theta_{\rm R}[i\omega ]
\end{array}
\right] \equiv \left[
\begin{array}{cc}
\Theta _{1}[i\omega ] & \Theta _{2}[i\omega ] \\
\Theta _{2}[i\omega ] & \Theta _{1}[i\omega ]
\end{array}
\right] .
\label{dec6_G}
\end{align}
The following lemma presents expressions for the quantities in Eqs. (\ref{3_key_a})-(\ref{3_key_d}).

\begin{lemma}
\label{lem:LL} In the limit $t_0 \to -\infty$, Eq. (\ref{3_key_a})-(\ref{3_key_d}) can be calculated by
\begin{subequations}
\begin{align}
&\langle 0_{\rm L}0_{\rm R}0\vert b_{\mathrm{out,L}}[i\omega
_{1}]b_{\mathrm{out,L}}[i\omega _{2}] b_{\rm L}^{\ast }[i\nu _{1}]b_{\rm R}^{\ast}[i\nu _{2}]\vert 0_{\rm L}0_{\rm R}0\rangle  \nonumber \\
=&\Theta _{2}[i\nu _{2}]\delta (\omega _{1}-\nu _{2})\delta
(\omega _{2}-\nu _{1}) \nonumber \\\
&+2\sqrt{\kappa }\Theta_{\rm L}[i\omega _{1}]\left[
\begin{array}{c}
f(\omega _{1},\omega _{2},\nu _{1},\nu _{2}) \\
f(\omega _{1},\omega _{2},\nu _{1},\nu _{2})
\end{array}
\right],
 \label{LL}
\\
\nonumber
\\
&\langle 0_{\rm L}0_{\rm R}0\vert b_{\mathrm{out,L}}[i\omega
_{1}]b_{\mathrm{out,R}}[i\omega _{2}]b_{\rm L}^{\ast }[i\nu _{1}]b_{\rm R}^{\ast}[i\nu _{2}]\vert 0_{\rm L}0_{\rm R}0\rangle  \nonumber \\
=&  \Theta _{1}[i\nu _{1}]\delta (\omega _{1}-\nu _{1})\delta
(\omega _{2}-\nu _{2})   \nonumber \\
&+2\sqrt{\kappa }\Theta_{\rm L}[i\omega _{1}]\left[
\begin{array}{c}
f(\omega _{1},\omega _{2},\nu _{1},\nu _{2}) \\
f(\omega _{1},\omega _{2},\nu _{1},\nu _{2})
\end{array}
\right],
 \label{LR}
\\
\nonumber
\\
& \langle 0_{\rm L}0_{\rm R}0\vert b_{\mathrm{out,R}}[i\omega_{1}]b_{\mathrm{out,R}}[i\omega _{2}]b_{\rm L}^{\ast }[i\nu _{1}]b_{\rm R}^{\ast}[i\nu _{2}]\vert 0_{\rm L}0_{\rm R}0\rangle
 \nonumber
\\
=&\Theta _{2}[i\nu _{1}]\delta (\omega _{1}-\nu _{1})\delta
(\omega _{2}-\nu _{2}) \nonumber\\
&+2\sqrt{\kappa }\Theta_{\rm R}[i\omega _{1}]\left[
\begin{array}{c}
f(\omega _{1},\omega _{2},\nu _{1},\nu _{2}) \\
f(\omega _{1},\omega _{2},\nu _{1},\nu _{2})
\end{array}
\right],
\label{RR}
\end{align}
\end{subequations}
where
\begin{align*}
f(\omega _{1},\omega _{2},\nu _{1},\nu _{2})  =&g(\omega _{1},\omega _{2},\nu _{1},\nu _{2})\delta (\nu _{1}+\nu
_{2}-\omega _{1}-\omega _{2})  \nonumber \\
&-\frac{\sqrt{\kappa }}{i(\omega _{2}+\omega _{c})+2\kappa }\delta (\omega
_{1}-\nu _{2})\delta (\nu _{1}-\omega _{2})\nonumber
\end{align*}
with
\begin{align}
&g(\omega _{1},\omega _{2},\nu _{1},\nu _{2})
\label{aug18_g}
 \\
\triangleq &-i\frac{\kappa ^{3/2}}{\pi }\frac{\nu _{1}+\nu _{2}+2\omega
_{c}-4i\kappa }{(\omega _{1}+\omega _{c}+2i\kappa )(\omega _{2}+\omega
_{c}-2i\kappa )}
\nonumber
\\
&  \times \frac{\nu _{1}+\nu _{2}+2\omega _{c}}{(\nu _{1}+\omega
_{c}-2i\kappa )(\nu _{2}+\omega _{c}-2i\kappa )(\nu _{1}+\nu _{2}+2\omega
_{c}-2i\kappa )}.  \nonumber
\end{align}
\end{lemma}

The proof of Lemma \ref{lem:LL} is given in the APPENDIX.

\begin{remark}
The Dirac delta function $\delta (\nu _{1}+\nu_{2}-\omega _{1}-\omega _{2})$ in $f(\omega _{1},\omega _{2},\nu _{1},\nu _{2})$ relates to the nonlinear frequency scattering of two photons. The output photons with frequencies $\omega_1$ and $\omega_2$ can be generated by any pair of incident photons with frequencies $\nu_1,\nu_2$ satisfying $\nu _{1}+\nu_{2}=\omega _{1}+\omega _{2}$. That is, the frequencies of the input photons may not be preserved.
\end{remark}

\begin{remark} \label{rem:freq}
Compared with the result for a two-photon single-qubit system \cite{Fan10}, there is an additional coefficient $(\nu _{1}+\nu _{2}+2\omega _{c})/(\nu _{1}+\nu _{2}+2\omega
_{c}-2i\kappa)$ in $g(\omega _{1},\omega _{2},\nu _{1},\nu _{2})$ that is associated with the nonlinear frequency scattering. Apparently, this term characterizes a two-photon process, where the two photons are taken as a single object with the frequency $\nu_1+\nu_2$ and interact with the combined two-qubit system with the detuning frequency $2\omega_c$. In particular, when the two photons are in resonance with the combined two-qubit system, i.e. $\nu_1+\nu_2+2\omega_c=0$, the nonlinear frequency scattering can be completely suppressed, which is impossible for a single-qubit system.
\end{remark}

On the basis of Lemma \ref{lem:LL} presented above, we are able to  derive the main result of this paper.

\begin{theorem}\label{thm:output_state}
The steady-state output two-photon state in Eq. (\ref{dec8_Psi}) can be calculated as
\begin{align}
& \ket{\Psi _{\mathrm{out}}}
\label{psi_out_ss}
 \\
=& \frac{1}{2}\int\limits_{\omega _{1},\omega _{2}} d\omega _{1} d\omega _{2} \, T_{\rm LL}[\omega _{1},\omega
_{2}]b_{\rm L}^{\ast }[i\omega _{1}]b_{\rm L}^{\ast }[i\omega _{2}]\ket{0_{\rm L}0_{\rm R}}
\nonumber
\\
& +\int\limits_{\omega _{1},\omega _{2}}d\omega _{1} d\omega _{2} \,T_{\rm LR}[\omega _{1},\omega
_{2}]b_{\rm L}^{\ast }[i\omega _{1}]b_{\rm R}^{\ast }[i\omega _{2}]\ket{0_{\rm L}0_{\rm R}}\nonumber
 \\
& +\frac{1}{2}\int\limits_{\omega _{1},\omega _{2}}d\omega _{1} d\omega _{2} \,T_{\rm RR}[\omega _{1},\omega
_{2}]b_{\rm R}^{\ast }[i\omega _{1}]b_{\rm R}^{\ast }[i\omega _{2}]\ket{0_{\rm L}0_{\rm R}},
\nonumber
\end{align}
where
\begin{subequations}
\begin{align}
&T_{\rm LL}[\omega _{1},\omega _{2}]
\label{T_LL2a}
\\
=&(1+T(\omega_1))S(\omega_2)\xi_{\rm L}[i\omega _{1}]\xi_{\rm R}[i\omega _{2}]+\chi(\omega_1,\omega_2),\nonumber
\\
\nonumber
\\
&T_{\rm LR}[\omega _{1},\omega _{2}]
\label{T_LR2a}
\\
=&\frac{(1+T(\omega_1))(1+T(\omega_2))}{4}\xi_{\rm L}[i\omega _{1}]\xi_{\rm R}[i\omega _{2}]  \nonumber \\
+&S(\omega_1)S(\omega_2)\xi_{\rm L}[i\omega _{2}]\xi_{\rm R}[i\omega _{1}]
+\chi(\omega_1,\omega_2),
\nonumber
\\
 \nonumber
\\
&T_{\rm RR}[\omega _{1},\omega _{2}]
\label{T_RR2a}
 \\
=&(1+T(\omega_2))S(\omega_1)\xi_{\rm L}[i\omega _{1}]\xi_{\rm R}[i\omega _{2}]+\chi(\omega_1,\omega_2)
  \nonumber
\end{align}
\end{subequations}
with
\begin{subequations}
\begin{align}
&T(\omega_i)\triangleq\frac{\omega _{i}+\omega _{c}+2i\kappa }{\omega
_{i}+\omega _{c}-2i\kappa },\ S(\omega_i)\triangleq\frac{2i\kappa}{\omega
_{i}+\omega _{c}-2i\kappa },
\label{eq:TS}
\\
\nonumber
\\
&\chi(\omega_1,\omega_2)\triangleq2\sqrt{\kappa }T(\omega_1)\int_{-\infty }^{\infty }d\nu _{1}\, \xi_{\rm L}[i\nu
_{1}]  \nonumber \\
&\times \xi_{\rm R}[i(\omega _{1}+\omega _{2}-\nu _{1})]g(\omega _{1},\omega
_{2},\nu _{1},\omega _{1}+\omega _{2}-\nu _{1}).
\label{eql:xi}
\end{align}
\end{subequations}
\end{theorem}

{\it Proof.}
Applying Lemma \ref{lem:LL} to Eq. (\ref{dec8_Psi}) proves the theorem. \hfill $\blacksquare$

\begin{remark}
It can be readily verified that the function $\chi(\omega_1,\omega_2)$ defined in Eq. (\ref{eql:xi}) satisfies $\chi(\omega_1,\omega_2)=\chi(\omega_2,\omega_1)$. Again, the nonlinear frequency scattering term $\chi(\omega_1,\omega_2)$ can be suppressed under the condition of two-photon resonance $\nu_1+\nu_2+2\omega_c=0$, cf. Remark \ref{rem:freq}. The physical meanings of the steady-state output field state $\ket{\Psi _{\mathrm{out}}}$ in Eq. \eqref{psi_out_ss} is clear:  it is a superposition state composed of three terms, which are two photons  in the left-going channel, one in each channel, and two in the right-going channel respectively.
\end{remark}

The following result presents a special case of Theorem \ref{thm:output_state}.

\begin{corollary} \label{cor:kappa}
Sending $\kappa \to 0$ while fixing all the other parameters, the steady-state output field state becomes
\begin{align}
& \left\vert \Psi _{\mathrm{out}}\right\rangle
\label{psi_out_ss_3}
\\
=&
\int_{-\infty}^\infty d\omega_1 \xi_{\rm L}[i\omega _{1}]b_{\rm L}^{\ast }[i\omega _{1}]|0_{\rm L}\rangle
\int_{-\infty}^\infty d\omega_2 \xi_{\rm R}[i\omega _{2}]b_{\rm R}^{\ast }[i\omega _{2}]|0_{\rm R}\rangle.
\nonumber
\end{align}
That is, the left-going output channel contains a single-photon packet $\xi_{\rm L}$, and  the right-going output channel contains a single-photon packet $\xi_{\rm R}$. On the other hand, Sending $\kappa \to \infty$ while fixing all the other parameters,  the steady-state output field state is
\begin{align}
& \left\vert \Psi _{\mathrm{out}}\right\rangle
\label{psi_out_ss_2}
\\
=&
\int_{-\infty}^\infty d\omega_1 \xi_{\rm R}[i\omega _{1}]b_{\rm L}^{\ast }[i\omega _{1}]|0_{\rm L}\rangle
\int_{-\infty}^\infty d\omega_2 \xi_{\rm L}[i\omega _{2}]b_{\rm R}^{\ast }[i\omega _{2}]|0_{\rm R}\rangle.
\nonumber
\end{align}
That is, the left-going output channel contains a single-photon packet $\xi_{\rm R}$, and  the right-going output channel contains a single-photon packet $\xi_{\rm L}$.
\end{corollary}

\begin{remark}
On one hand, when the coupling strength $\kappa$ is small, the interaction between the two-level systems and the input photons is weak. In the limit $\kappa \to 0$,  the left- (right-) going photon will be in the left (right) output channel. This interprets in Eq. (\ref{psi_out_ss_3}). On the other hand, in the strong coupling limit $\kappa \to \infty$, each two-level system acts as a mirror so that each input photon is bounced back. This interprets Eq. (\ref{psi_out_ss_2}).
\end{remark}

\subsection{The probabilities}\label{subsec:prob}

Let $P_{\rm LL}$ denote the probability of finding two photons in the left-going output channel $b_{\rm out,L}$, $P_{\rm RR}$ the probability of finding two photons in the  right-going output channel $b_{\rm out,R}$, and $P_{\rm LR}$  the probability of finding one photon in each output channel, respectively. By Theorem \ref{thm:output_state}, we have
\begin{eqnarray*}
&&P_{\rm LL}  = \frac{1}{4}\int\limits_{\omega _{1},\omega _{2}}\left\vert T_{\rm LL}[\omega _{1},\omega _{2}]\right\vert ^{2}+T_{\rm LL}^{\ast }[\omega _{1},\omega _{2}]T_{\rm LL}[\omega _{2},\omega _{1}],
\\
&&P_{\rm LR} = \int\limits_{\omega _{1},\omega _{2}}\left\vert
T_{\rm LR}[\omega _{1},\omega _{2}]\right\vert ^{2},
\\
&&P_{\rm RR} = \frac{1}{4}\int\limits_{\omega _{1},\omega _{2}}\left\vert T_{\rm RR}[\omega _{1},\omega _{2}]\right\vert ^{2}+T_{\rm RR}^{\ast }[\omega _{1},\omega _{2}]T_{\rm RR}[\omega _{2},\omega _{1}].
\end{eqnarray*}
In particular, when $\xi_{\rm L}\equiv \xi_{\rm R}$, we get
\begin{subequations}
\begin{eqnarray}
P_{\rm LL} =P_{\rm RR}  &=& \frac{1}{2}\int\limits_{\omega _{1},\omega _{2}}\left\vert
T_{\rm LL}[\omega _{1},\omega _{2}]\right\vert ^{2},  \label{aug26_LL_RR}\\
P_{\rm LR} &=&\int\limits_{\omega _{1},\omega _{2}}\left\vert T_{\rm LR}[\omega
_{1},\omega _{2}]\right\vert ^{2}. \label{aug26_LR}
\end{eqnarray}
\end{subequations}

\section{Synthesis of systems using the coherent feedback structure}\label{sec:app}

\subsection{Tunable Hong-Ou-Mandel (HOM) interferometer}\label{subsec:HOM}

\begin{figure}[tbp]
\centering
\includegraphics[scale=0.3]{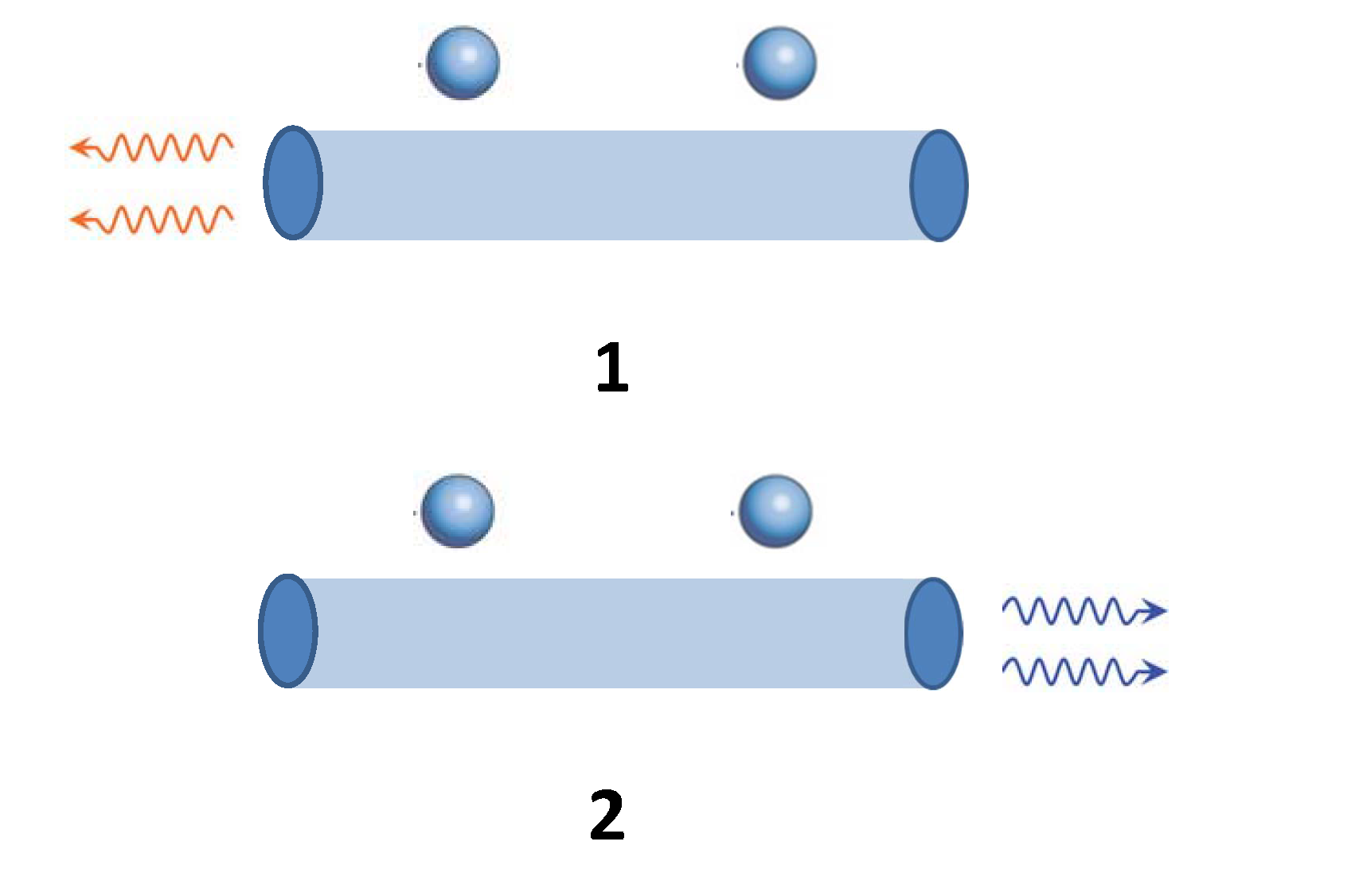}
\caption{The two continuous-mode output photons are either left-going or right-going simultaneously, with equal probability. There is no possibility of finding one photon in an output channel and the other in another output channel.}
\label{fig_hom}
\end{figure}

The HOM effect refers to a two-photon interference effect that occurs when two identical photons enter a balanced beam splitter, one in each input port \cite{HOM87}. In our case, the left-going and right-going input photons enter the system in different ports; Fig. \ref{fig_cf}. In the classical HOM experiment, due to the destructive interference, the two output photons appear in the same output port, with equal probability. In our case, it means that the two photons simultaneously leave the network from either the left- or right-going channel with equal probability; Fig. \ref{fig_hom}.

By Corollary \ref{cor:kappa},  if $\kappa\to0$ or $\kappa\to\infty$ while all the other parameters are fixed,  in the steady state there will be one photon in each output channel. In this subsection, we show that controlling the detuning $\omega_c$ can turn the coherent feedback network into a tunable  HOM interferometer.   Notice that the detuning $\omega_c$ is indeed physically controllable using artificial qubits \cite{Neumeier13}, \cite{ZB13}, \cite{Laasko14}.

\begin{figure}[!htb]
\centering
\begin{minipage}{6.0cm}
\centering
\includegraphics[width=1.2\textwidth]{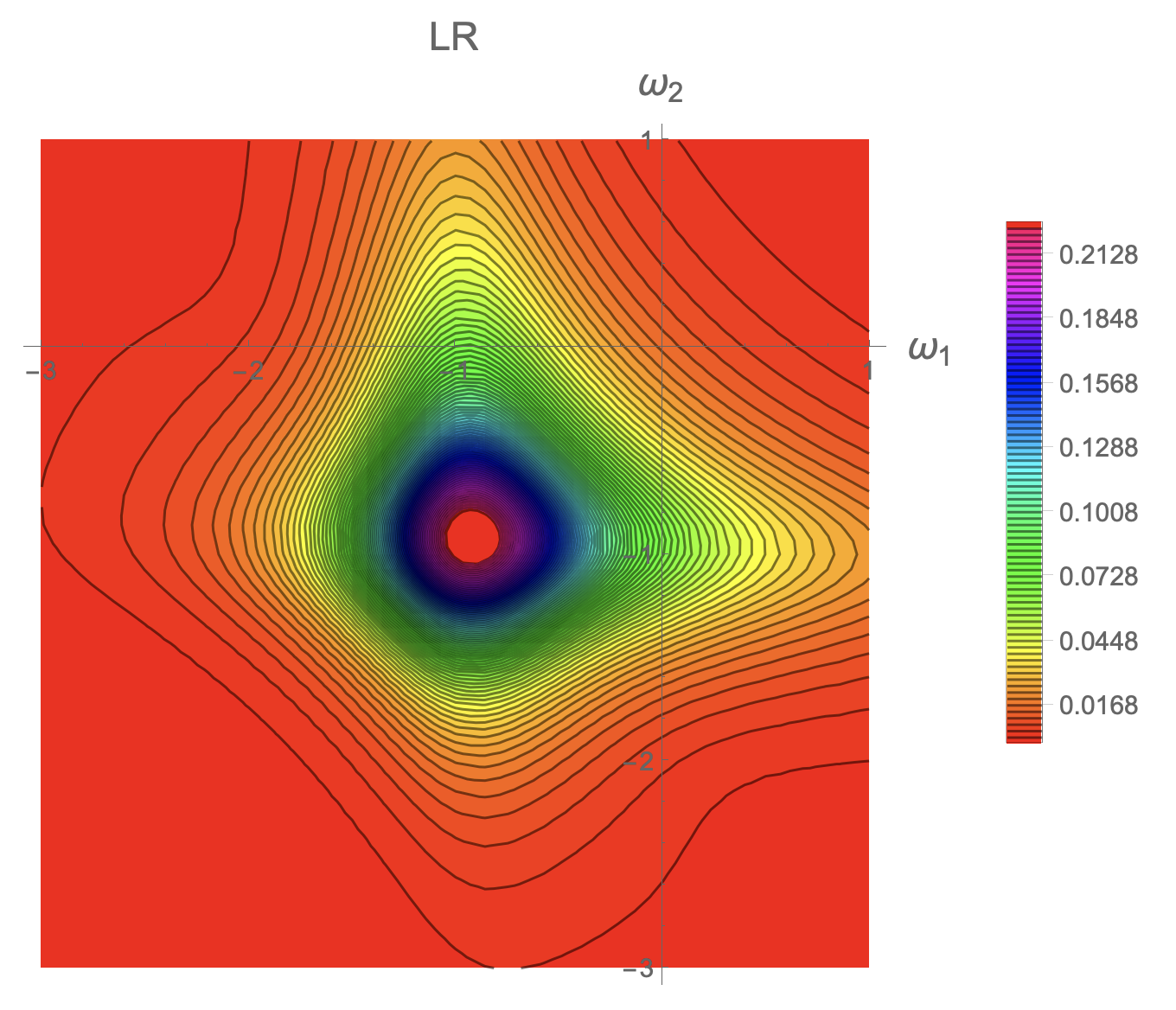}
\end{minipage}
\begin{minipage}{6.0cm}
\centering
\includegraphics[width=1.2\textwidth]{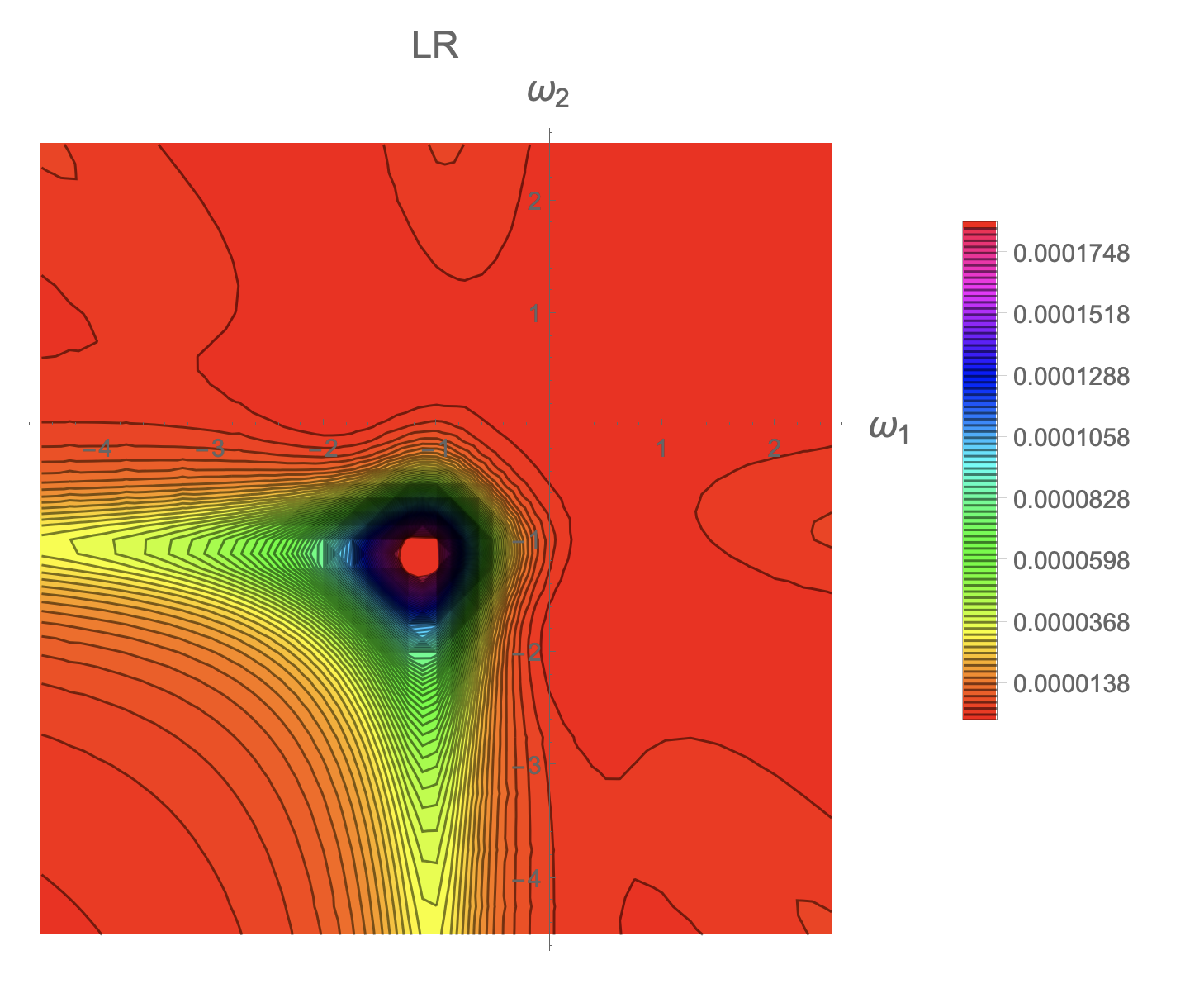}
\end{minipage}
\caption{\label{fig_jan27} $|T_{\rm LR}(\omega_1,\omega_2)|^2$ with parameters $\gamma=1, \omega_o=1,\kappa=1.5$,  $\omega_c=0$ (for the upper subfigure), and $\omega_c=3$ (for the lower subfigure).}
\end{figure}

Assume $\xi_{\rm L}=\xi_{\rm R}\equiv \xi$. Let $\omega_c = \zeta \kappa$ for some $\zeta\geq0$.
According to Eqs. \eqref{aug18_2a} and \eqref{eql:xi}, for any given $\omega_1, \omega_2\in \mbb{R}$, $\lim_{\kappa \to \infty}  \chi(\omega_1,\omega_2) = 0$.
As a  result
\begin{align*}
&\lim_{\kappa \to \infty} T_{\rm LR}[\omega _{1},\omega _{2}]
\nonumber
\\
=& \lim_{\kappa \to \infty}\frac{(\omega_1+\zeta\kappa)(\omega_2+\zeta\kappa)- 4\kappa^2}{(\omega_1+\zeta\kappa-2i\kappa)(\omega_2+\zeta\kappa-2i\kappa)}\xi[i\omega_1]\xi[i\omega_2]
\nonumber
\\
=&
\frac{\zeta^2 - 4}{(\zeta-2i)^2} \xi[i\omega_1]\xi[i\omega_2].
\end{align*}
Thus, when $\zeta = 2$, $\lim_{\kappa \to \infty} T_{\rm LR}[\omega _{1},\omega _{2}]=0$. Then, by Eq. \eqref{aug26_LR}, $\lim_{\kappa \to \infty}  P_{\rm LR}=0$. In a similar way, it can be shown that $\lim_{\kappa \to \infty}  P_{\rm LL}=\lim_{\kappa \to \infty}  P_{\rm RR} = \frac{1}{2}$. That is, two photons {\it simultaneously} appear in the left- or right-going channel with equal probability. This is the famous HOM interference phenomenon. Let us look at an example. In Fig. \ref{fig_jan27},  the identical input photons have a wave-packet of Lorentzian type with FWHM $\gamma=1$ and carrier frequency $\omega_o=1$. In the upper subfigure the detuning frequency $\omega_c=0$, while in the lower subfigure the detuning frequency $\omega_c=3=2\kappa$. It can be seen clearly that in the  lower subfigure $T_{\rm LR}[\omega _{1},\omega _{2}] $ is very close to zero. Indeed, when $\omega_c=2\kappa$, numerical simulations show that $P_{\rm LR}\to 0$ as $\kappa\to \infty$.

In that follows we demonstrate that the continuous-mode setting discussed in this paper is also applicable to the single-mode setting. Let the two single-photon input states be those in Example \ref{ex:photon}. In this case, $\xi_{\rm L}[i\nu]=\xi_{\rm R}[i\nu]=\xi[i\nu]$. Notice that
\begin{equation}\label{eq:aug28_delta}
\lim_{\gamma\rightarrow0}\left|\xi[i\nu]\right|^2 =  \lim_{\gamma\rightarrow0}\frac{1}{\pi} \frac{\gamma/2}{(\nu-\omega_o)^2+(\gamma/2)^2} = \delta(\nu-\omega_o).
\end{equation}
In other words, in the limit $\gamma\rightarrow0$, the inputs turn to monochromatic lights, i.e. photons with single frequency $\omega_o$.  It is easy to verify that
$\lim_{\kappa \to 0}\chi(\omega_1,\omega_2)=0$, where $\chi(\omega_1,\omega_2)$ is defined in Eq. \eqref{eql:xi}. Now consider $\omega_o+\omega_c=\beta\kappa$ in which $\omega_c$ is no longer a fixed value but dependent on $\kappa$. By Eq. \eqref{eq:aug28_delta}, it can be shown that
\begin{eqnarray*}
&&\lim_{\gamma \to 0}\int\limits_{\omega _{1},\omega _{2}}|(1+T(\omega_1))S(\omega_2)\xi[i\omega _{1}]\xi[i\omega _{2}]|^2\nonumber\\
&=&|(1+T(\omega_o))S(\omega_o)|^2
=
\left(\frac{4\beta}{\beta^2+4}\right)^2.
\end{eqnarray*}
As a result, when $\beta = 2$,  by Eq. \eqref{aug26_LL_RR},
$
\lim_{\gamma, \kappa\to0} P_{\rm LL} = P_{\rm RR}  =\frac{1}{2}.
$
Similarly, it can be shown that
$
\lim_{\gamma, \kappa\to0} P_{\rm LR} =0.
$
In other words, the two photons simultaneously leave the network from either the left- or right-going channel with equal probability.

\subsection{Marginally stable single-photon device}\label{sec:1_photon}
As shown in \cite{YJ14}, \cite{PZJ15}, the interaction between a two-level system and a single photon can be fully analyzed using a transfer function approach. Similarly, if the coherent feedback network in Fig. \ref{fig_sys} has a single-photon input (e.g., the left-going input field  $b_{\rm L}$ contains a single photon while the right-going input field $b_{\rm R}$ is in the vacuum state), then essentially the network dynamics can be investigated by means of linear systems theory. Unfortunately,  as the matrix $A$ in Eq. \eqref{A} is not Hurwitz stable, the linear transfer function approach in \cite{YJ14}, \cite{PZJ15} is not applicable. However, the general input-output analysis presented in Section \ref{sec:main_result} indeed works in the single-photon case.

Let us assume that  the left-going input field $b_{\rm L}$ is still in the single-photon state  $b_{\rm L}^{\ast }(\xi_{\rm L})\vert 0_{\rm L}\rangle$, and the right-going input field $b_{\rm R}$ is in the vacuum state $|0_{\rm R}\rangle$.
Then,  the joint system-field state is
\begin{equation*}
\ket{\Psi (t)}=U(t, t_{0})b_{\rm L}^{\ast }(\xi
_{1})\ket{0_{\rm L}0_{\rm R}0}.  \label{aug28_infty_t}
\end{equation*}
In the steady-state case ($t_{0}\rightarrow -\infty ,t\rightarrow \infty $), the single photon leaves the feedback-connected two-level
systems in their ground state. As a result, the steady-state output single-photon
state is
\begin{eqnarray}
\ket{\Psi _{\mathrm{out}}} &=&\lim_{t_{0}\rightarrow
-\infty ,t\rightarrow \infty }\braket{0|\Psi (t)}
\nonumber\\
&=&\lim_{t_{0}\rightarrow -\infty ,t\rightarrow \infty }\left\langle
0\right\vert U(t, t_{0})b_{\rm L}^{\ast }(\xi_{\rm L})\left\vert
0_{\rm L}0_{\rm R}0\right\rangle.  \label{sout}
\end{eqnarray}
With the time-domain 1-photon basis for the input field given by
\begin{equation*}
\left\{ \int_{-\infty}^{\infty }dp_{1}\ket{1_{{\rm L}p_{1}}}
\bra{1_{{\rm L}p_{1}}},\ \int_{-\infty}^{\infty }dp_{1}\
\ket{1_{{\rm R}p_{1}}} \bra{1_{{\rm R}p_{1}}}
\right\},  \label{aug28_dec6:basis}
\end{equation*}
Eq. (\ref{sout}) can be simplified as
\begin{align}
& \ket{\Psi _{\mathrm{out}}}
\nonumber
  \\
=&\lim_{t_0\to-\infty} \int_{-\infty}^\infty dp_{1}\ket{1_{{\rm L}p_{1}}} \int_{t_{0}}^{\infty}dt_{1} \xi_{\rm L}(t_{1})
\nonumber
\\
\times&\left\langle 0_{\rm L}0_{\rm R}0\right\vert b_{
\mathrm{out,L}}(p_{1})b_{\rm L}^{\ast }(t_{1})\left\vert
0_{\rm L}0_{\rm R}0\right\rangle
 \nonumber
  \\
+&\lim_{t_0\to-\infty} \int_{-\infty}^\infty dp_{1}\ket{1_{{\rm R}p_{1}}} \int_{t_{0}}^{\infty
}dt_{1}\ \xi_{\rm L}(t_{1})
\nonumber
\\
\times& \ \ \  \left\langle 0_{\rm L}0_{\rm R}0\right\vert b_{
\mathrm{out,R}}(p_{1})b_{\rm L}^{\ast }(t_{1})\left\vert
0_{\rm L}0_{\rm R}0\right\rangle .
 \label{aug28_mar17_1}
\end{align}
As with the two-photon case, we go to the frequency domain by applying the Fourier
transform to the time variables $t_{1}$ and $p_{1}$, respectively. In the frequency domain, Eq. (\ref{aug28_mar17_1}) becomes
\begin{align}
& \ket{\Psi _{\mathrm{out}}}  \label{aug28_dec8_Psi} \\
=&\int\limits_{\omega_1,\nu_1}\xi_{\rm L}[i\nu _{1}]\ket{1_{\rm L\omega _{1}}}\left\langle 0_{\rm L}0_{\rm R}0\right\vert b_{\mathrm{out,L}}[i\omega
_{1}]b_{\rm L}^{\ast }[i\nu _{1}]\ket{0_{\rm L}0_{\rm R}0}
\nonumber \\
+&\int\limits_{\omega_1,\nu_1}\xi_{\rm L}[i\nu _{1}]\ket{1_{\rm R\omega _{1}}}\left\langle 0_{\rm L}0_{\rm R}0\right\vert b_{\mathrm{out,R}}[i\omega
_{1}]b_{\rm L}^{\ast }[i\nu _{1}]\ket{0_{\rm L}0_{\rm R}0}.
\nonumber
\end{align}
Therefore, we have to calculate the following quantities:
\begin{subequations}
\begin{align}
& \left\langle 0_{\rm L}0_{\rm R}0\right\vert b_{\mathrm{out,L}}[i\omega
_{1}]b_{\rm L}^{\ast }[i\nu _{1}]\left\vert 0_{\rm L}0_{\rm R}0\right\rangle ,
\label{aug28_3_key_a} \\
& \left\langle 0_{\rm L}0_{\rm R}0\right\vert b_{\mathrm{out,R}}[i\omega
_{1}]b_{\rm L}^{\ast }[i\nu _{1}]\left\vert 0_{\rm L}0_{\rm R}0\right\rangle .
\label{aug28_3_key_d}
\end{align}
\end{subequations}
First, we consider Eq. (\ref{aug28_3_key_a}). By Eqs. (\ref{sys_f}) and (\ref{dec20_10}) in the APPENDIX we have
\begin{align}
& \left\langle 0_{\rm L}0_{\rm R}0\right\vert b_{\mathrm{out},\rm L}(p_{1})b_{\rm L}^{\ast }[i\nu _{1}]\left\vert
0_{\rm L}0_{\rm R}0\right\rangle  \label{aug28_eq:jun4_2}\\
=& \frac{1}{\sqrt{2\pi }}\int\limits_{\omega _{1}}e^{i\omega
_{1}p_{1}}\left\langle 0_{\rm L}0_{\rm R}0\right\vert\Theta_{\rm L}[i\omega
_{1}]b_{\mathrm{in}}[i\omega _{1}]b_{\rm L}^{\ast }[i\nu _{1}]\left\vert
0_{\rm L}0_{\rm R}0\right\rangle.\nonumber
\end{align}
Using (\ref{sept9_1}) in the APPENDIX, in the limit $t_{0}\rightarrow -\infty $, Eq. (\ref{aug28_eq:jun4_2}) can be simplified to be
\begin{align}
&\frac{1}{\sqrt{2\pi }}  \int\limits_{\omega _{1}}e^{i\omega _{1}p_{1}}\left\langle
0_{\rm L}0_{\rm R}0\right\vert
\nonumber
 \\
&\times (\Theta _{1}[i\omega _{1}]b_{\rm L}[i\omega _{1}]+\Theta _{2}[i\omega
_{1}]b_{\rm R}[i\omega _{1}])b_{\rm L}^{\ast }[i\nu _{1}]
\left\vert0_{\rm L}0_{\rm R}0\right\rangle  
\nonumber
\\
=&
\frac{1}{\sqrt{2\pi }}\Theta _{1}[i\nu _{1}]e^{i\nu _{1}p_{1}}.
 \label{aug28_july1_2}
\end{align}
By Eqs. (\ref{aug28_eq:jun4_2})-(\ref{aug28_july1_2}), we have
\begin{equation}\label{single_LL}
\left\langle 0_{\rm L}0_{\rm R}0\right\vert b_{\mathrm{out,L}}[i\omega
_{1}]b_{\rm L}^{\ast }[i\nu _{1}]\left\vert 0_{\rm L}0_{\rm R}0\right\rangle
=
\Theta _{1}[i\nu _{1}]\delta (\omega _{1}-\nu _{1}).
\end{equation}
Eq. (\ref{aug28_3_key_d}) can be calculated via a similar way as
\begin{eqnarray}
&&\left\langle 0_{\rm L}0_{\rm R}0\right\vert b_{\mathrm{out,R}}[i\omega
_{1}]b_{\rm L}^{\ast }[i\nu _{1}]\left\vert 0_{\rm L}0_{\rm R}0\right\rangle
\nonumber
\\
&=&\frac{1}{\sqrt{2\pi }}\int\limits_{p_{1}}e^{-i\omega
_{1}p_{1}}\frac{1}{\sqrt{2\pi }}\Theta _{2}[i\nu _{1}]e^{i\nu _{1}p_{1}}
\nonumber
\\
&=&\Theta _{2}[i\nu _{1}]\delta (\omega _{1}-\nu _{1}).
\label{single_LR}
\end{eqnarray}
Substituting Eqs. (\ref{single_LL}) and (\ref{single_LR}) into Eq. (\ref{aug28_dec8_Psi}) yields the steady-state output single-photon state, which is
\begin{align}
& \ket{\Psi _{\mathrm{out}}}
\nonumber
 \\
=&  \int\limits_\omega\xi_{\rm L}[i\omega ]\bigg( \frac{1+T(\omega)}{2}b_{\rm L}^{\ast }[i\omega ]+S(\omega)b_{\rm R}^{\ast }[i\omega ]\bigg)  \ket{0_{\rm L}0_{\rm R}}
 \nonumber
 \\
=&\int\limits_\omega\left( G[i\omega ]\left[
\begin{array}{c}
\xi_{\rm L}[i\omega ] \\
0
\end{array}
\right] \right) ^{T}b_{\mathrm{in}}^{\#}[i\omega ] \ket{0_{\rm L}0_{\rm R}}.
 \label{aug28_Psi_out}
\end{align}
Denote
\begin{eqnarray*}
\eta_{\rm L}[i\omega ]&=&\frac{\omega +\omega _{c}}{\omega +\omega
_{c}-2i\kappa }\xi_{\rm L}[i\omega ]=\frac{1+T(\omega)}{2}\xi_{\rm L}[i\omega ], \\
\eta_{\rm R}[i\omega ]&=&\frac{2i\kappa }{\omega +\omega _{c}-2i\kappa }\xi_{\rm L}[i\omega ]=S(\omega)\xi_{\rm L}[i\omega ],
\label{eta_R}
\end{eqnarray*}
and substitute them into Eq. (\ref{aug28_Psi_out}) yields
\begin{equation}
\ket{\Psi _{\mathrm{out}}} =\int\limits_\omega\left( \eta
_{L}[i\omega ]b_{\rm L}^{\ast }[i\omega ]+\eta_{\rm R}[i\omega ]b_{\rm R}^{\ast
}[i\omega ]\right) \ket{0_{\rm L}0_{\rm R}}.
  \label{sept10_1}
\end{equation}
$\ket{\Psi _{\mathrm{out}}}$ is normalized since $|\eta
_{L}[i\omega ]|^2+|\eta_{\rm R}[i\omega ]|=|\xi_{\rm L}[i\omega ]|^2$. Clearly,
\begin{subequations}
\begin{align}
\lim_{\kappa \rightarrow 0 }\left\vert \Psi _{\mathrm{out}}\right\rangle
=&
 \int_{-\infty}^\infty d\omega \ \xi_{\rm L}[i\omega ]b_{\rm L}^{\ast }[i\omega ]|0_{\rm L}\rangle \otimes |0_{\rm R}\rangle.
\label{jan_91}
\\
\lim_{\kappa \rightarrow \infty }\left\vert \Psi _{\mathrm{out}}\right\rangle
=&
|0_{\rm L}\rangle \otimes  \int_{-\infty}^\infty d\omega \ (-\xi_{\rm L}[i\omega ])b_{\rm R}^{\ast }[i\omega ]|0_{\rm R}\rangle,
\label{sept10_2b}
\end{align}
\end{subequations}
which are consistent with Eqs. (\ref{psi_out_ss_3})-(\ref{psi_out_ss_2}).

\begin{remark}
According to Eq. (\ref{aug28_Psi_out}), the pulse shape of  the photon in the output channels is obtained by linearly transforming that of the input photon by $G[i\omega]$.  This looks like a linear dynamics. Indeed, as shown in \cite{YJ14,PZJ15}, the interaction between a two-level system and a single photon can be fully analyzed in a transfer function approach. Unfortunately,  as the coherent feedback network studied in this paper is only marginally stable, the linear transfer function approach in \cite{YJ14,PZJ15} is not applicable. However, as shown above, the general framework presented Section \ref{sec:main_result} indeed works.
\end{remark}

\begin{remark}\label{rem:single}
 It is worthwhile to notice that Eq. (\ref{sept10_2b}) is consistent with   \cite[Fig. 3]{ZGB10} for single-photon Fock-state scattering. That is, for strong coupling, a two-level atom appears as a mirror so that the input single photon is reflected.  This is true even with the existence of a nonzero detuning $\omega_c$.
\end{remark}

\section{Conclusion} \label{sec:conclusion}
In this paper, we have studied a coherent feedback network which consists of two identical qubits and is marginally stable. The coherent feedback network can be physically realized by integrating a two-qubit system with one-dimensional waveguide, which is suitable for applications in nano-photonic quantum networks and information processing on-chip. Due to the feedback loop, the two input photons can be confined between the qubits with a probability, leading to multiple times of photon-photon interaction and enhanced nonlinearity. The previous works \cite{ZB13,Laasko14,YH15} have not analytically solved for the steady-state system response when the network is only marginally stable and modeled by Markovian QSDEs. In this paper, we introduce the input-output formalism and study the system response in the steady-state limit \cite{ZJ13,Fan10} which fully captures the time-correlation of the output photons. More importantly, by Theorem \ref{thm:output_state} we have provided an end-to-end solution that exactly describes the input-output relation for two generic continuous-mode photons.

A novel two-photon process has been found in the nonlinear response of this coherent feedback network, which provides additional options for controlling the nonlinearity. In particular, under a condition of two-photon resonance the nonlinear frequency scattering can be completely suppressed, which is never possible for two photons that interact via a single qubit. The coherent feedback system is readily integrable with the existing nanophotonic circuitry. Since one- and two-photon operations are sufficient for universal optical quantum computing, the method of this paper is easily scalable to practical-sized quantum information processing circuits.

{\bf Acknowledgements.}   The authors wish to thank Mazyar Mirrahimi for his very helpful discussions.


\noindent {\bf Appendix.}

\setcounter{lemma}{0}
\renewcommand{\thelemma}{A.\arabic{lemma}}

In order to prove Lemma \ref{lem:LL}, we need to establish Lemma \ref{lem:key}.

For $i=1,2$, define functions
\begin{subequations}
\begin{align}
&f_{{\rm L},i}(\omega _{1},\omega _{2},\nu _{1},\nu _{2})  \label{eq:jun4_5c} \\
\triangleq&\left\langle 0_{\rm L}0_{\rm R}0\right\vert b_{\rm L}[i\omega _{1}]\sigma
_{-,i}[i\omega _{2}]b_{\rm L}^{\ast }[i\nu _{1}]b_{\rm R}^{\ast }[i\nu
_{2}]\left\vert 0_{\rm L}0_{\rm R}0\right\rangle ,
 \nonumber
\\
{\rm and}
 \nonumber
\\
& f_{{\rm R},i}(\omega _{1},\omega _{2},\nu _{1},\nu _{2})  \label{eq:jun4_5d} \\
\triangleq&\left\langle 0_{\rm L}0_{\rm R}0\right\vert b_{\rm R}[i\omega _{1}]\sigma
_{-,i}[i\omega _{2}]b_{\rm L}^{\ast }[i\nu _{1}]b_{\rm R}^{\ast }[i\nu
_{2}]\left\vert 0_{\rm L}0_{\rm R}0\right\rangle , \nonumber
\end{align}
\end{subequations}
respectively.

\begin{lemma}\label{lem:key}
The functions defined in Eqs. (\ref{eq:jun4_5c})-(\ref{eq:jun4_5d}) satisfy
\begin{eqnarray}
&&f_{{\rm L},1}(\omega _{1},\omega _{2},\nu _{1},\nu _{2}) = f_{{\rm L},2}(\omega _{1},\omega _{2},\nu _{1},\nu _{2})   \nonumber\\
&=&f_{{\rm R},1}(\omega _{1},\omega _{2},\nu _{1},\nu _{2}) = f_{{\rm R},2}(\omega _{1},\omega _{2},\nu _{1},\nu _{2})  \nonumber \\
&=&g(\omega _{1},\omega _{2},\nu _{1},\nu _{2})\delta (\nu _{1}+\nu
_{2}-\omega _{1}-\omega _{2})  \nonumber \\
&&-\frac{\sqrt{\kappa }}{i(\omega _{2}+\omega _{c})+2\kappa }\delta (\omega
_{1}-\nu _{2})\delta (\nu _{1}-\omega _{2}).  \label{aug18_2a}
\end{eqnarray}
\end{lemma}

{\it Proof of Lemma \ref{lem:key}.} For the matrix $A$ defined in Eq. (\ref{A}), we have its matrix exponential
\begin{equation}
e^{A}=e^{-i\omega _{c}-\kappa }\left[
\begin{array}{cc}
\cosh \kappa & -\sinh \kappa \\
-\sinh \kappa & \cosh \kappa
\end{array}
\right].  \label{exp_A}
\end{equation}
Thus,
\begin{equation}
e^{At}C=e^{-(i\omega _{c}+2\kappa )t}C.  \label{dec20_1}
\end{equation}
 Substituting Eq. (\ref{dec20_1}) into Eq. (\ref{sol_1}), together with the commutation relations in Eq. \eqref{CCR_t},   we get
\begin{eqnarray}
&&\hspace{-2mm}\langle 0_{\rm L}0_{\rm R}0\vert \left[
\begin{array}{c}
\sigma _{-,1}(t) \\
\sigma _{-,2}(t)
\end{array}
\right] b_{\rm L}^{\ast }(r)\left\vert 0_{\rm L}0_{\rm R}0\right\rangle
\nonumber
 \\
&=&\langle 0_{\rm L}0_{\rm R}0\vert \left[
\begin{array}{c}
\sigma _{-,1}(t) \\
\sigma _{-,2}(t)
\end{array}
\right] b_{\rm R}^{\ast }(r)\left\vert 0_{\rm L}0_{\rm R}0\right\rangle  \nonumber
\\
&=&-\sqrt{\kappa }\int_{t_{0}}^{t}d\tau \ e^{-(i\omega _{c}+2\kappa )(t-\tau
)}\delta (\tau -r)\left[
\begin{array}{c}
1 \\
1
\end{array}
\right].  \label{feb26_1}
\end{eqnarray}
Applying the Fourier transform to Eq. (\ref{feb26_1}) with respect to the time variable $r$ yields
\begin{align}
&\left\langle 0_{\rm L}0_{\rm R}0\right\vert \left[
\begin{array}{c}
\sigma _{-,1}(t) \\
\sigma _{-,2}(t)
\end{array}
\right] b_{\rm L}^{\ast }[i\omega ]\left\vert 0_{\rm L}0_{\rm R}0\right\rangle
\nonumber
 \\
=&\left\langle 0_{\rm L}0_{\rm R}0\right\vert \left[
\begin{array}{c}
\sigma _{-,1}(t) \\
\sigma _{-,2}(t)
\end{array}
\right] b_{\rm R}^{\ast }[i\omega ]\left\vert 0_{\rm L}0_{\rm R}0\right\rangle
\nonumber \\
=&i\sqrt{\frac{\kappa }{2\pi }}e^{i\omega t}\frac{1-e^{-(2\kappa +i(\omega
_{c}+\omega ))(t-t_{0})}}{(\omega _{c}+\omega )-2i\kappa }\left[
\begin{array}{c}
1 \\
1
\end{array}
\right].
\label{temp2}
\end{align}
Adjoining both sides of Eq. (\ref{temp2}) and by Eqs. (\ref{b_s_2})-(\ref{CCR_t}), it is straightforward to show that
\begin{eqnarray}
&&\left\langle 0_{\rm L}0_{\rm R}0\right\vert b_{\rm L}(t)b_{\rm L}^{\ast }[i\omega
]\left\vert 0_{\rm L}0_{\rm R}0\right\rangle
\nonumber
\\
&=&
\left\langle 0_{\rm L}0_{\rm R}0\right\vert b_{\rm R}(t)b_{\rm R}^{\ast }[i\omega
]\left\vert 0_{\rm L}0_{\rm R}0\right\rangle
\nonumber
\\
&=&
\frac{1}{\sqrt{2\pi }}e^{i\omega t}, \ \ {\rm as} \ \ t_0\to -\infty.
\label{sept9_1}
\end{eqnarray}
By means of Eqs. (\ref{sys_e}) and the fact that $\sigma _{z}=2\sigma _{+}\sigma _{-}-I$,  differentiating the vector functions $\left[
\begin{array}{c}
f_{{\rm L},1}(\omega _{1},p_{2},\nu _{1},\nu _{2}) \\
f_{{\rm L},2}(\omega _{1},p_{2},\nu _{1},\nu _{2})
\end{array}
\right] $ with respect to the time variable $p_{2}$ yields
\begin{subequations}
\begin{align}
& \frac{\partial }{\partial p_{2}}\left[
\begin{array}{c}
f_{{\rm L},1}(\omega _{1},p_{2},\nu _{1},\nu _{2}) \\
f_{{\rm L},2}(\omega _{1},p_{2},\nu _{1},\nu _{2})
\end{array}
\right]
 \nonumber
 \\
=& A\left[
\begin{array}{c}
f_{{\rm L},1}(\omega _{1},p_{2},\nu _{1},\nu _{2}) \\
f_{{\rm L},2}(\omega _{1},p_{2},\nu _{1},\nu _{2})
\end{array}
\right]
 \nonumber
  \\
& +2\kappa \left\langle 0_{\rm L}0_{\rm R}0\right\vert b_{\rm L}[i\omega _{1}]
\left[
\begin{array}{c}
\sigma _{+,1}(p_{2}) \\
\sigma _{+,2}(p_{2})
\end{array}
\right] \left\vert 0_{\rm L}0_{\rm R}0\right\rangle
 \nonumber
  \\
& \times \left\langle 0_{\rm L}0_{\rm R}0\right\vert \sigma
_{-,1}(p_{2})\sigma _{-,2}(p_{2})b_{\rm L}^{\ast }[i\nu _{1}]b_{\rm R}^{\ast }[i\nu
_{2}]\left\vert 0_{\rm L}0_{\rm R}0\right\rangle
 \nonumber
  \\
& +2\sqrt{\kappa }\left\langle 0_{\rm L}0_{\rm R}0\right\vert b_{\rm L}[i\omega
_{1}]
\nonumber
  \\
& \times
\left[
\begin{array}{cc}
\sigma _{+,1}(p_{2}) & 0 \\
0 & \sigma _{+,2}(p_{2})
\end{array}
\right] \left\vert 0_{\rm L}0_{\rm R}0\right\rangle
\nonumber
 \\
& \ \ \ \ \times \left\langle 0_{\rm L}0_{\rm R}0\right\vert \left[
\begin{array}{c}
\sigma _{-,1}(p_{2}) \\
\sigma _{-,2}(p_{2})
\end{array}
\right] \left( b_{\rm L}(p_{2})+b_{\rm R}(p_{2})\right)
\nonumber
\\
& \ \ \ \ \times b_{\rm L}^{\ast }[i\nu _{1}]b_{\rm R}^{\ast }[i\nu _{2}]\left\vert
0_{\rm L}0_{\rm R}0\right\rangle
 \nonumber
  \\
& -\sqrt{\kappa }\left[
\begin{array}{c}
1 \\
1
\end{array}
\right] \left\langle 0_{\rm L}0_{\rm R}0\right\vert b_{\rm L}[i\omega
_{1}]\left( b_{\rm L}(p_{2})+b_{\rm R}(p_{2})\right)
 \nonumber
 \\
& \ \ \ \ \times b_{\rm L}^{\ast }[i\nu _{1}]b_{\rm R}^{\ast }[i\nu _{2}]\left\vert
0_{\rm L}0_{\rm R}0\right\rangle .
 \label{july1_5f}
\end{align}
\end{subequations}
The non-homogeneous terms of the ODEs (\ref{july1_5f}) can be calculated using Eqs. (\ref{temp2}) and (\ref{sept9_1}) except
\begin{align}
\Pi _{ij}(t)
\triangleq  \left\langle 0_{\rm L}0_{\rm R}0\right\vert \sigma
_{-,i}(t)\sigma _{-,j}(t)b_{\rm L}^{\ast }[i\nu _{1}]b_{\rm R}^{\ast }[i\nu
_{2}]\left\vert 0_{\rm L}0_{\rm R}0\right\rangle
\label{Pi_ij}
\end{align}
whose initial condition are
\begin{align}
&\Pi _{ij}(t_{0})
\label{Pi_initial}
\\
=&\left\langle 0_{\rm L}0_{\rm R}0\right\vert \sigma
_{-,i}(t_{0})\sigma _{-,j}(t_{0})b_{\rm L}^{\ast }[i\nu _{1}]b_{\rm R}^{\ast }[i\nu
_{2}]\left\vert 0_{\rm L}0_{\rm R}0\right\rangle
\nonumber
\\
=&0.\nonumber
\end{align}
Re-write Eq. (\ref{sys_e}) in the It\^{o} form,
\begin{eqnarray*}
d\sigma _{-,1}(t) &=&\alpha \sigma _{-,1}(t)dt+\kappa \sigma _{z,1}(t)\sigma
_{-,2}(t)dt \\
&&+\sqrt{\kappa }\sigma _{z,1}(t)\left( dB_{\rm L}(t)+dB_{\rm R}(t)\right) , \\
d\sigma _{-,2}(t) &=&\alpha \sigma _{-,2}(t)dt+\kappa \sigma _{z,2}(t)\sigma
_{-,1}(t)dt \\
&&+\sqrt{\kappa }\sigma _{z,2}(t)\left( dB_{\rm L}(t)+dB_{\rm R}(t)\right),
\end{eqnarray*}
where $dB_{\rm j}(t) \equiv \int_{t}^{t+dt} b_{\rm j}(\tau)d\tau$  are It\^{o} increments, ($j={\rm L}, {\rm R}$). By It\^{o} calculus we have
\begin{eqnarray*}
&&d(\sigma _{-,1}(t)\sigma _{-,1}(t)) \\
&=&\alpha \sigma _{-,1}(t)\sigma _{-,1}(t)dt+\kappa \sigma _{z,1}(t)\sigma
_{-,1}(t)\sigma _{-,2}(t)dt \\
&&+\sqrt{\kappa }\sigma _{z,1}(t)\sigma _{-,1}(t)\left(
dB_{\rm L}(t)+dB_{\rm R}(t)\right)  \\
&&+\alpha \sigma _{-,1}(t)\sigma _{-,1}(t)dt+\kappa \sigma _{-,1}(t)\sigma
_{-,2}(t)\sigma _{z,1}(t)dt \\
&&+\sqrt{\kappa }\sigma _{-,1}(t)\sigma _{z,1}(t)\left(
dB_{\rm L}(t)+dB_{\rm R}(t)\right)  \\
&&+\kappa \sigma _{z,1}(t)^{2}\left( dB_{\rm L}(t)+dB_{\rm R}(t)\right) ^{2} \\
&=&2\alpha \sigma _{-,1}(t)\sigma _{-,1}(t),
\end{eqnarray*}
where the fact
\begin{eqnarray*}
\sigma _{-,1}(t)\sigma _{z,1}(t) = \sigma _{-,1}(t), \ \
\sigma _{z,1}(t)\sigma _{-,1}(t) = -\sigma _{-,1}(t)
\end{eqnarray*}
have been used to derive the last step. Therefore $d\Pi _{11} = 2\alpha \Pi _{11}dt$, which, under the initial condition (\ref{Pi_initial}), has the trivial solution
\begin{equation}
\Pi _{11}(t)\equiv 0,\ \ t\geq t_{0}.  \label{PI_11}
\end{equation}
Similarly, it can be shown that
\begin{equation}
\Pi _{22}(t)\equiv 0,\ \ t\geq t_{0}.  \label{PI_22}
\end{equation}
Next, we look at $\Pi _{12}(t)$ (which equals $\Pi _{21}(t)$). By It\^{o} calculus,
\begin{eqnarray}
&&d(\sigma _{-,1}(t)\sigma _{-,2}(t))  \label{aug3_d} \\
&=&\alpha \sigma _{-,1}(t)\sigma _{-,2}(t)dt+\kappa \sigma _{z,1}(t)\sigma
_{-,2}^{2}(t)dt  \nonumber \\
&&+\sqrt{\kappa }\sigma _{z,1}(t)\sigma _{-,2}(t)\left(
dB_{\rm L}(t)+dB_{\rm R}(t)\right)   \nonumber \\
&&+\alpha \sigma _{-,1}(t)\sigma _{-,2}(t)dt+\kappa \sigma _{z,2}(t)\sigma
_{-,1}^{2}(t)dt  \nonumber \\
&&+\sqrt{\kappa }\sigma _{z,2}(t)\sigma _{-,1}(t)\left(
dB_{\rm L}(t)+dB_{\rm R}(t)\right)   \nonumber \\
&&+\kappa \sigma _{z,1}(t)\sigma _{z,2}(t)\left( dB_{\rm L}(t)+dB_{\rm R}(t)\right)
^{2}.  \nonumber
\end{eqnarray}
Noticing
\begin{align*}
&\left\langle 0_{\rm L}0_{\rm R}0\right\vert \sigma _{z,1}(t)\sigma
_{z,2}(t)\left( dB_{\rm L}(t)+dB_{\rm R}(t)\right) ^{2} \\
&\times b_{\rm L}^{\ast }[i\nu _{1}]b_{\rm R}^{\ast }[i\nu _{2}]\left\vert
0_{\rm L}0_{\rm R}0\right\rangle  \\
=&\left\langle 0_{\rm L}0_{\rm R}0\right\vert dB_{\rm L}(t)^{2}b_{\rm L}^{\ast
}[i\nu _{1}]b_{\rm R}^{\ast }[i\nu _{2}]\left\vert
0_{\rm L}0_{\rm R}0\right\rangle  \\
&+\left\langle 0_{\rm L}0_{\rm R}0\right\vert dB_{\rm R}(t)^{2}b_{\rm L}^{\ast
}[i\nu _{1}]b_{\rm R}^{\ast }[i\nu _{2}]\left\vert
0_{\rm L}0_{\rm R}0\right\rangle  \\
&+2\left\langle 0_{\rm L}0_{\rm R}0\right\vert
dB_{\rm L}(t)dB_{\rm R}(t)b_{\rm L}^{\ast }[i\nu _{1}]b_{\rm R}^{\ast }[i\nu _{2}]\left\vert
0_{\rm L}0_{\rm R}0\right\rangle  \\
=&0,
\end{align*}
Eq. (\ref{aug3_d}) yields
\begin{eqnarray}
&&d\Pi _{12}
\label{d_II}
 \\
&=&2\alpha \Pi _{12}dt-\frac{\sqrt{\kappa }}{\sqrt{2\pi }}e^{i\nu
_{1}t}\left\langle 0_{\rm L}0_{\rm R}0\right\vert (\sigma _{-,1}(t)+\sigma
_{-,2}(t))  \nonumber \\
&&\times b_{\rm R}^{\ast }[i\nu _{2}]\left\vert
0_{\rm L}0_{\rm R}0\right\rangle dt  \nonumber \\
&-&\frac{\sqrt{\kappa }}{\sqrt{2\pi }}e^{i\nu _{2}t}\left\langle
0_{\rm L}0_{\rm R}0\right\vert (\sigma _{-,1}(t)+\sigma _{-,2}(t))b_{\rm L}^{\ast }[i\nu _{1}]\left\vert
0_{\rm L}0_{\rm R}0\right\rangle dt.\nonumber
\end{eqnarray}
Moreover, by Eq. (\ref{temp2}) we can explicitly calculate the non-homogeneous terms of Eq. (\ref{d_II}) and get
\begin{align}
&d\Pi _{12}  \nonumber \\
=&2\alpha \Pi _{12}dt-2i\frac{\kappa }{2\pi }e^{i(\nu _{1}+\nu _{2})t}\frac{
1-e^{-(2\kappa +i(\omega _{c}+\nu _{2}))(t-t_{0})}}{(\omega _{c}+\nu
_{2})-2i\kappa }dt  \nonumber \\
&-2i\frac{\kappa }{2\pi }e^{i(\nu _{1}+\nu _{2})t}\frac{1-e^{-(2\kappa
+i(\omega _{c}+\nu _{1}))(t-t_{0})}}{(\omega _{c}+\nu _{1})-2i\kappa }dt,
\nonumber
\end{align}
which in the limit $\ t_{0}\rightarrow -\infty $ reduces to
\begin{align}
& \dot{\Pi}_{12}
\nonumber
\\
=&2\alpha \Pi _{12}
\nonumber
\\
& -\frac{i\kappa }{\pi }\left( \frac{1}{(\omega _{c}+\nu _{2})-2i\kappa }+
\frac{1}{(\omega _{c}+\nu _{1})-2i\kappa }\right) e^{i(\nu _{1}+\nu _{2})t},
\nonumber
\end{align}
whose solution is
\begin{eqnarray}
&&\Pi _{12}(t)
 \nonumber
 \\
&=&-\frac{i\kappa }{\pi }\frac{2\omega _{c}+\nu _{1}+\nu _{2}-4i\kappa }{
((\omega _{c}+\nu _{2})-2i\kappa )((\omega _{c}+\nu _{1})-2i\kappa )}
 \nonumber
\\
&&\times \int_{t_{0}}^{t}e^{2\alpha (t-r)}e^{i(\nu _{1}+\nu _{2})r}  \nonumber \\
&\rightarrow &-\mathrm{i}\frac{\kappa }{\pi }\frac{2\omega _{c}+\nu _{1}+\nu
_{2}-4i\kappa }{((\omega _{c}+\nu _{2})-2i\kappa )((\omega _{c}+\nu
_{1})-2i\kappa )}
 \nonumber
\\
&&\times
\frac{e^{i(\nu _{1}+\nu _{2})t}}{i(\nu _{1}+\nu
_{2})-2\alpha },\ {\rm as}\ \ t_{0}\rightarrow -\infty
\label{PI12}.
\end{eqnarray}
As a result, the Fourier transform of the ODEs (\ref{july1_5f})  with respect to the time variable $p_2$ can be written as
\begin{subequations}
\begin{eqnarray}
&&i\omega _{2}\left[
\begin{array}{c}
f_{{\rm L},1}(\omega _{1},\omega _{2},\nu _{1},\nu _{2}) \\
f_{{\rm L},2}(\omega _{1},\omega _{2},\nu _{1},\nu _{2})
\end{array}
\right]
\nonumber
 \\
&=&A\left[
\begin{array}{c}
f_{{\rm L},1}(\omega _{1},\omega _{2},\nu _{1},\nu _{2}) \\
f_{{\rm L},2}(\omega _{1},\omega_{2},\nu _{1},\nu _{2})
\end{array}
\right]   \nonumber \\
&&-\frac{2\kappa ^{5/2}}{\pi }\frac{2\omega _{c}+\nu _{1}+\nu _{2}-4i\kappa
}{((\omega _{c}+\nu _{2})-2i\kappa )((\omega _{c}+\nu _{1})-2i\kappa )}
\nonumber \\
&&\times \frac{\delta (\nu _{1}+\nu _{2}-\omega _{1}-\omega _{2})}{((\omega
_{c}+\omega _{1})+2i\kappa )(i(\nu _{1}+\nu _{2})-2\alpha )}\left[
\begin{array}{c}
1 \\
1
\end{array}
\right]   \nonumber \\
&&+\frac{\kappa ^{3/2}}{\pi }\frac{2\omega _{c}+\nu _{1}+\nu _{2}-4i\kappa }{
((\omega _{c}+\nu _{2})-2i\kappa )((\omega _{c}+\nu _{1})-2i\kappa )}
\nonumber \\
&&\times \frac{\delta (\nu _{1}+\nu _{2}-\omega _{1}-\omega _{2})}{(\omega
_{c}+\omega _{1})+2i\kappa }\left[
\begin{array}{c}
1 \\
1
\end{array}
\right]   \nonumber \\
&&-\sqrt{\kappa }\delta (\omega _{1}-\nu _{1})\delta (\nu _{2}-\omega _{2})
\left[
\begin{array}{c}
1 \\
1
\end{array}
\right].
\label{aug6_f_L}
\end{eqnarray}
\end{subequations}
The expression of $\left[
\begin{array}{c}
f_{{\rm L},1}(\omega _{1},p_{2},\nu _{1},\nu _{2}) \\
f_{{\rm L},2}(\omega _{1},p_{2},\nu _{1},\nu _{2})
\end{array}
\right]$ is then derived using Eq. (\ref{aug6_f_L}). The expression of $\left[
\begin{array}{c}
f_{{\rm R},1}(\omega _{1},p_{2},\nu _{1},\nu _{2}) \\
f_{{\rm R},2}(\omega _{1},p_{2},\nu _{1},\nu _{2})
\end{array}
\right]$ can be obtained similarly, which completes the proof. \hfill $\blacksquare$

{\it Proof of Lemma \ref{lem:LL}.}  By Lemma \ref{lem:key}, we can write $f_{{\rm L},1}(\omega _{1},\omega _{2},\nu _{1},\nu _{2}) = f_{{\rm L},2}(\omega _{1},\omega _{2},\nu _{1},\nu _{2})=f_{{\rm R},1}(\omega _{1},\omega _{2},\nu _{1},\nu _{2}) = f_{{\rm R},2}(\omega _{1},\omega _{2},\nu _{1},\nu _{2}) \equiv f(\omega _{1},\omega _{2},\nu _{1},\nu _{2})$. Fourier transforming both sides of  Eq. (\ref{sol_1}) with respect to the time  variable $t$ yields
\begin{align}
&  \left\langle 0_{\rm L}0_{\rm R}0\right\vert \left[
\begin{array}{c}
\sigma _{-,1}[i\omega ] \\
\sigma _{-,2}[i\omega ]
\end{array}
\right]  \label{dec19_1} \\
=& \frac{1}{\sqrt{2\pi }}\int_{t_{0}}^{\infty }dt\ e^{-(i\omega
-A)t}e^{-At_{0}}\left\langle 0_{\rm L}0_{\rm R}0\right\vert \left[
\begin{array}{c}
\sigma _{-,1}(t_{0}) \\
\sigma _{-,2}(t_{0})
\end{array}
\right]  \nonumber \\
& -\frac{\sqrt{\kappa }}{\sqrt{2\pi }}\int_{t_{0}}^{\infty }dt\ e^{-i\omega t}\int_{t_{0}}^{t}dr\ e^{A(t-r)}C\left\langle0_{\rm L}0_{\rm R}0\right\vert b_{\mathrm{in}}(r).
\nonumber
\end{align}
First, we look at the second term on the right-hand side of Eq. (\ref{dec19_1}). By
Eqs. (\ref{b_t_1}) and  (\ref{dec20_1}), we get
\begin{eqnarray}
&&\frac{\sqrt{\kappa }}{\sqrt{2\pi }}\int_{t_{0}}^{\infty }dt\ \
e^{-i\omega t}\int_{t_{0}}^{t}dr\ e^{A(t-r)}C\left\langle
0_{\rm L}0_{\rm R}0\right\vert b_{\mathrm{in}}(r)  \nonumber \\
&=&\frac{\sqrt{\kappa }}{2\kappa +i(\omega _{c}+\omega )}C\left\langle
0_{\rm L}0_{\rm R}0\right\vert b_{\mathrm{in}}[i\omega ].\label{dec20_3}
\end{eqnarray}
Next, we look at the first term on the right-hand side of Eq. (\ref{dec19_1}). Performing eigen-structure decomposition on $i\omega -A$ gives
\begin{eqnarray*}
i\omega -A &=&V\left[
\begin{array}{cc}
i(\omega +\omega _{c}) & 0 \\
0 & 2\kappa +i(\omega +\omega _{c})
\end{array}
\right] V,  \nonumber \\
\end{eqnarray*}
where the columns of the matrix
\begin{equation*}
V\triangleq \frac{1}{\sqrt{2}}\left[
\begin{array}{cc}
-1 & 1 \\
1 & 1
\end{array}
\right]
\end{equation*}
are eigenvectors of the matrix $i\omega -A$. Then
\begin{align}
\int_{t_{0}}^{\infty }dt\ e^{-(i\omega -A)t}
&=\frac{e^{-(2\kappa +i(\omega +\omega _{c}))t_{0}}}{2\left( 2\kappa
+i(\omega +\omega _{c})\right) }C  \nonumber \\
&+\frac{\pi }{2}\delta (\omega +\omega _{c})e^{-i(\omega +\omega _{c})t_{0}}
\left[
\begin{array}{cc}
1 & -1 \\
-1 & 1
\end{array}
\right].  \label{dec20_6}
\end{align}
Thus, the solution of Eq. (\ref{dec19_1}) is obtained as
\begin{eqnarray}
&&\bra{0_{\rm L}0_{\rm R}0} \left[
\begin{array}{c}
\sigma _{-,1}[i\omega ] \\
\sigma _{-,2}[i\omega ]
\end{array}
\right]
\label{dec20_9}
\\
&=&\frac{1}{\sqrt{2\pi }}\frac{e^{-(2\kappa +i(\omega +\omega _{c}))t_{0}}}{%
2\left( 2\kappa +i(\omega +\omega _{c})\right) }Ce^{-At_{0}}
\nonumber
\\
&& \ \ \ \times
\bra{
0_{\rm L}0_{\rm R}0} \left[
\begin{array}{c}
\sigma _{-,1}(t_{0}) \\
\sigma _{-,2}(t_{0})
\end{array}
\right]  \nonumber \\
&&+\frac{1}{\sqrt{2\pi }}\frac{\pi }{2}\delta (\omega +\omega
_{c})e^{-i(\omega +\omega _{c})t_{0}}\left[
\begin{array}{cc}
1 & -1 \\
-1 & 1
\end{array}
\right]  \nonumber \\
&& \ \ \ \times e^{-At_{0}}\bra{0_{\rm L}0_{\rm R}0}\left[
\begin{array}{c}
\sigma _{-,1}(t_{0}) \\
\sigma _{-,2}(t_{0})
\end{array}
\right]  \nonumber \\
&&-\frac{\sqrt{\kappa }}{2\kappa +i(\omega _{c}+\omega )}C\bra{
0_{\rm L}0_{\rm R}0} b_{\mathrm{in}}[i\omega ].\nonumber
\end{eqnarray}
Similarly, we can establish the following equation.
\begin{eqnarray}
&&\bra{0_{\rm L}0_{\rm R}0} \left[
\begin{array}{c}
b_{\mathrm{out,L}}[i\omega ] \\
b_{\mathrm{out,R}}[i\omega ]
\end{array}
\right]
\label{dec20_10}\\
&=&\frac{\sqrt{\kappa }}{\sqrt{2\pi }}\frac{e^{-i\omega t_{0}}}{2\kappa
+i(\omega +\omega _{c})}C\bra{0_{\rm L}0_{\rm R}0}\left[
\begin{array}{c}
\sigma _{-,1}(t_{0}) \\
\sigma _{-,2}(t_{0})
\end{array}
\right]  \nonumber \\
&&+G[i\omega ]\bra{0_{\rm L}0_{\rm R}0} b_{\mathrm{in}
}[i\omega ].
\nonumber
\end{eqnarray}
By Eq. (\ref{sys_f}) and Eq. (\ref{dec20_10}), we have
\begin{subequations}
\begin{align}
& \left\langle 0_{\rm L}0_{\rm R}0\right\vert b_{\mathrm{out,L}}(p_{1})b_{
\mathrm{out,L}}(p_{2})b_{\rm L}^{\ast }[i\nu _{1}]b_{\rm R}^{\ast }[i\nu
_{2}]\left\vert 0_{\rm L}0_{\rm R}0\right\rangle
 \label{july8_5}
 \\
& =\frac{1}{\sqrt{2\pi }}\int_{-\infty }^{\infty }d\omega _{1}\ e^{i\omega
_{1}p_{1}}\left\langle 0_{\rm L}0_{\rm R}0\right\vert
\nonumber
  \\
& \ \ \ \ \ \times b_{\mathrm{out,\rm L}}[i\omega _{1}]b_{\mathrm{out
,L}}(p_{2})b_{\rm L}^{\ast }[i\nu _{1}]b_{\rm R}^{\ast }[i\nu _{2}]\left\vert
0_{\rm L}0_{\rm R}0\right\rangle
\nonumber
\\
& =\frac{\kappa }{2\pi }\int_{-\infty }^{\infty }d\omega _{1}\ \frac{
e^{i\omega _{1}(p_{1}-t_{0})}}{2\kappa +i(\omega _{1}+\omega _{c})}
\nonumber
\\
&  \ \ \ \ \ \times
\left\langle 0_{\rm L}0_{\rm R}0\right\vert (\sigma _{-,1}(t
_{0})+\sigma _{-,2}(t_{0}))
\nonumber
 \\
& \ \  \ \times [ 1 \ \ \ 1] \left[
\begin{array}{c}
\sigma _{-,1}(p_{2}) \\
\sigma _{-,2}(p_{2})
\end{array}
\right] b_{\rm L}^{\ast }[i\nu _{1}]b_{\rm R}^{\ast }[i\nu _{2}]\left\vert
0_{\rm L}0_{\rm R}0\right\rangle
\label{dec22_2b}
 \\
& +\frac{\sqrt{\kappa }}{2\pi }\int_{-\infty }^{\infty }d\omega _{1}\ \frac{
e^{i\omega _{1}(p_{1}-t_{0})}}{2\kappa +i(\omega _{1}+\omega _{c})}
\nonumber
\\
& \ \ \ \ \ \times
\left\langle 0_{\rm L}0_{\rm R}0\right\vert (\sigma _{-,1}(t
_{0})+\sigma _{-,2}(t_{0}))
 \nonumber
 \\
& \ \ \ \ \ \times b_{\rm L}(p_{2})b_{\rm L}^{\ast }[i\nu _{1}]b_{\rm R}^{\ast }[i\nu
_{2}]\left\vert 0_{\rm L}0_{\rm R}0\right\rangle
\label{dec22_2c}
\\
& +\frac{\sqrt{\kappa }}{\sqrt{2\pi }}\int_{-\infty }^{\infty }d\omega _{1}\
e^{i\omega _{1}p_{1}}\Theta_{\rm L}[i\omega _{1}]\left\langle
0_{\rm L}0_{\rm R}0\right\vert
\nonumber
\\
& \ \ \ \ \times b_{\mathrm{in}}[i\omega _{1}]\left[
\begin{array}{cc}
1 & 1
\end{array}
\right] \left[
\begin{array}{c}
\sigma _{-,1}(p_{2}) \\
\sigma _{-,2}(p_{2})
\end{array}
\right]
\nonumber
\\
& \ \ \ \ \times  b_{\rm L}^{\ast }[i\nu _{1}]b_{\rm R}^{\ast }[i\nu _{2}]\left\vert
0_{\rm L}0_{\rm R}0\right\rangle
\label{eq:jun4_4}
\\
& +\frac{1}{\sqrt{2\pi }}\int_{-\infty }^{\infty }d\omega _{1}\ e^{i\omega
_{1}p_{1}}\left\langle 0_{\rm L}0_{\rm R}0\right\vert\Theta_{\rm L}[i\omega _{1}]
\nonumber
 \\
& \ \ \ \ \times b_{\mathrm{in}}[i\omega
_{1}]b_{\rm L}(p_{2})b_{\rm L}^{\ast }[i\nu _{1}]b_{\rm R}^{\ast }[i\nu _{2}]\left\vert
0_{\rm L}0_{\rm R}0\right\rangle .
 \label{eq:jun4_2}
\end{align}
\end{subequations}
Firstly, notice
\begin{align*}
&\left\langle 0_{\rm L}0_{\rm R}0\right\vert (\sigma _{-,1}(t
_{0})+\sigma _{-,2}(t_{0}))
\\
& \ \ \ \ \times b_{\rm L}(p_{2})b_{\rm L}^{\ast }[i\nu
_{1}]b_{\rm R}^{\ast }[i\nu _{2}]\left\vert 0_{\rm L}0_{\rm R}0\right\rangle =0.
\end{align*}
Secondly, by (\ref{sept9_1}), in the limit $t_{0}\rightarrow -\infty
$, Eq. (\ref{eq:jun4_2}) can be simplified to be
\begin{eqnarray}
&&\frac{1}{\sqrt{2\pi }} \int_{-\infty}^\infty d\omega _{1}\ e^{i\omega _{1}p_{1}}\Theta
_{2}[i\omega _{1}]\left\langle 0_{\rm L}0_{\rm R}0\right\vert
\nonumber \\
&&\times \delta (\omega
_{1}-\nu _{2})\delta (\omega _{2}-\nu _{1})  \nonumber \\
&=&\frac{1}{2\pi }\Theta _{2}[i\nu _{2}]e^{i\nu _{2}p_{1}}e^{i\nu _{1}p_{2}}.
\label{july1_2}
\end{eqnarray}
Thirdly, denote
\begin{eqnarray}
&& h_{ij}(p_{2},\nu _{1},\nu _{2}) \nonumber
\\
 &\triangleq& \left\langle
0_{\rm L}0_{\rm R}0\right\vert \sigma _{-,i}(t_{0})\sigma
_{-,j}(p_{2})b_{\rm L}^{\ast }[i\nu _{1}]b_{\rm R}^{\ast }[i\nu _{2}]\left\vert
0_{\rm L}0_{\rm R}0\right\rangle,\nonumber\\ \label{dec22_3b}
\end{eqnarray}
and substitute Eq. (\ref{dec22_3b}) into Eq. (\ref{dec22_2b}) yields
\begin{eqnarray}
&&\sum_{i,j=1}^{2}\frac{\kappa }{2\pi }\int_{-\infty }^{\infty
}d\omega _{1}\ \frac{e^{i\omega _{1}(p_{1}-t_{0})}}{2\kappa +i(\omega
_{1}+\omega _{c})}h_{ij}(p_{2},\nu _{1},\nu _{2})  \nonumber \\
&=&\kappa \sqrt{2\pi }e^{-(2\kappa +i\omega
_{c})(p_{1}-t_{0})}\sum_{i,j=1}^{2}h_{ij}(p_{2},\nu _{1},\nu _{2}),
\label{june6_2b}
\end{eqnarray}
where the fact
\begin{equation}
\int_{-\infty }^{\infty }\frac{e^{i\omega t}}{i\omega +(2\kappa +i\omega
_{c})}d\omega =\left\{
\begin{array}{cc}
2\pi e^{-(2\kappa +i\omega _{c})t}, & t\geq 0, \\
0, & t<0
\end{array}
\right.  \label{july2_delta2}
\end{equation}
has been used. Since $\lim_{t_{0}\rightarrow -\infty }e^{-(2\kappa +i\omega _{c})(p_{1}-t_{0})}=0$, Eq. (\ref{dec22_2b}) goes to $0$ as $t_{0}\rightarrow -\infty$. Finally, Eq. (\ref
{eq:jun4_4}) can be written as
\begin{equation}
\frac{2\sqrt{\kappa }}{\sqrt{2\pi }}\int_{-\infty }^{\infty }d\omega _{1}\
e^{i\omega _{1}p_{1}}\Theta_{\rm L}[i\omega _{1}]\left[
\begin{array}{c}
f(\omega _{1},p_{2},\nu _{1},\nu _{2})\\
f(\omega _{1},p_{2},\nu _{1},\nu _{2})
\end{array}
\right].\label{june6_2}
\end{equation}
Consequently, in the limit $t_{0}\rightarrow -\infty $, Eq. (\ref{july8_5}) becomes
\begin{align}
&\left\langle 0_{\rm L}0_{\rm R}0\right\vert b_{\mathrm{out,L}}(p_{1})b_{
\mathrm{out,L}}(p_{2})b_{\rm L}^{\ast }[i\nu _{1}]b_{\rm R}^{\ast }[i\nu
_{2}]\left\vert 0_{\rm L}0_{\rm R}0\right\rangle
\label{aug11_1}
  \\
=&\frac{1}{2\pi }\Theta _{2}[i\nu _{2}]e^{i\nu _{2}p_{1}}e^{i\nu _{1}p_{2}}
\nonumber \\
&+\frac{2\sqrt{\kappa }}{\sqrt{2\pi }}\int_{-\infty }^{\infty }d\omega
_{1}\ e^{i\omega _{1}p_{1}}\Theta_{\rm L}[i\omega _{1}]\left[
\begin{array}{c}
f(\omega _{1},p_{2},\nu _{1},\nu _{2}) \\
f(\omega _{1},p_{2},\nu _{1},\nu _{2})
\end{array}
\right] .  \nonumber
\end{align}
By applying the Fourier transform to Eq. (\ref{aug11_1})  we obtain Eq. (\ref{LL}).  Eqs. (\ref{LR})-(\ref{RR}) can be established in a similar way. \hfill $\blacksquare$

\bibliographystyle{IFAC}

\begin{thebibliography}{99}

\bibitem{AA03}
F. Albertini and D. D'Alessandro.
\newblock Notions of controllability for bilinear multilevel quantum systems.
\newblock {\em IEEE Trans. Automat. Contr.}, 48:1399--1403, 2003.

\bibitem{AT12}
C. Altafini and F. Ticozzi.
\newblock Modeling and control of quantum systems: an introduction.
\newblock {\em IEEE Trans. Automat. Contr.}, 57:1898--1917, 2012.



\bibitem{BCS09}
B. Bonnard, M. Chyba, and D. Sugny.
\newblock Time-minimal control of dissipative two-level quantum systems: the
  generic case.
\newblock {\em IEEE Trans. Automat. Contr.}, 54:2598--2610, 2009.


\bibitem{BrodA16}
D. J. Brod, J. Combes and J. Gea-Banacloche.
\newblock{Two photons co- and counterpropagating through $N$ cross-Kerr sites}.
\newblock{\em Phys. Rev. A}, 94:023833, 2016.

\bibitem{Brod16}
D. J. Brod and J. Combes.
\newblock {Passive CPHASE gate via cross-Kerr nonlinearities}.
\newblock {\em Phys. Rev. Lett.}, 117:080502, 2016.


\bibitem{CKS17}
J. Combes, J. Kerckhoff, and M. Sarovar.
\newblock The SLH framework for modeling quantum input-output networks.
\newblock {\em Advances in Physics: X}, 2(3):784--888, 2017.

\bibitem{TMT15}
T. Caneva, M. T. Manzoni, T. Shi, J. S. Douglas, J. I. Cirac, and D. E. Chang.
\newblock Quantum dynamics of propagating photons with strong interactions: a generalized input-output formalism.
\newblock {\em New Journal of Physics}, 17:113001, 2015.

\bibitem{DZA19}
Z. Dong, G. Zhang,  and N. H. Amini.
\newblock On the response of a two-level system to two-photon inputs.
\newblock  {\em SIAM Journal on Control and Optimization}, 57(5): 3445-3470, 2019.


\bibitem{DZA19b}
Z. Dong, G. Zhang, and Nina H. Amini.
\newblock Quantum filtering for a two-level atom driven by two counter-propagating photons.
\newblock  {\em Quantum information Processing}, 18:136, 2019.


\bibitem{DP10}
D. Dong and I. R. Petersen.
\newblock Quantum control theory and applications:a survey.
\newblock {\em IET Control Theory \& Applications}, 4:2651--2671, 2010.

\bibitem{SVF18}
S. Das, V. E. Elfving, F. Reiter, and A. S. Sørensen.
\newblock Photon scattering from a system of multilevel quantum emitters. I. Formalism.
\newblock {\em Phys. Rev. A}, 91:043837, 2018.

\bibitem{Fan10}
S. Fan, S. E. Kocabas, and J. T. Shen.
\newblock Input-output formalism for few-photon transport in one-dimensional
  nanophotonic waveguides coupled to a qubit.
\newblock {\em Phys. Rev. A.}, 82:063821, 2010.

\bibitem{YH15}
Y. L. Fang, H. U. Baranger.
\newblock Waveguide QED: Power spectra and correlations of two photons scattered off multiple distant qubits and a mirror.
\newblock {\em Phys. Rev. A}, 91:053845, 2015.


\bibitem{GC85}
C. W. Gardiner and M. J. Collett.
\newblock Input and output in damped quantum systems: Quantum stochastic
  differential equations and the master equation.
\newblock {\em Phys. Rev. A}, 31:3761--3774, Jun 1985.

\bibitem{GZ00}
C. W. Gardiner and P. Zoller.
\newblock {\em Quantum Noise}.
\newblock Springer, 2000.

\bibitem{GJ09}
J. E. Gough and M. R. James.
\newblock The series product and its application to quantum feedforward and
  feedback networks.
\newblock {\em IEEE Trans. Automat. Contr.}, 54:2530--2544, 2009.

\bibitem{GHN+12}
J. E. Gough, M. R James, H. I. Nurdin, and J. Combes.
\newblock Quantum filtering for systems driven by fields in single-photon
  states or superposition of coherent states.
\newblock {\em Phys. Rev. A.}, 86(4):043819, 2012.


\bibitem{HOM87}
C. K. Hong, Z. Y. Ou, and L. Mandel.
\newblock Measurement of subpicosecond time intervals between two photons by
  interference.
\newblock {\em Phys. Rev. Lett.}, 59:2044, 1987.

\bibitem{HP84}
R. L. Hudson and K. R. Parthasarathy.
\newblock Quantum It\^o's formula and stochastic evolutions.
\newblock {\em Communications in Mathematical Physics}, 93(3):301--323, Sep
  1984.

\bibitem{JNP08}
M. R. James, H. I. Nurdin, and I. R. Petersen.
\newblock $H^{\infty}$ control of linear quantum stochastic systems.
\newblock {\em IEEE Trans. Automat. Control}, 53:1787--1803, 2008.

\bibitem{kimble08}
H. Kimble.
\newblock The quantum internet.
\newblock {\em Nature}, 453(7198):1023--1030, 2008.

\bibitem{Kolchin11}
P. Kolchin, R. F. Oulton, and X. Zhang.
\newblock Nonlinear quantum optics in a waveguide: distinct single photons
  strongly interacting at the single atom level.
\newblock {\em Phys. Rev. Lett.}, 106:113601, 2011.

\bibitem{K16}
S. E. Kocabaş.
\newblock Few-photon scattering in dispersive waveguides with multiple qubits.
\newblock {\em Opt. Lett.}, 41:2533, 2016.

\bibitem{WJ17}
W. Konyk and J. Gea-Banacloche.
\newblock One- and two-photon scattering by two atoms in a waveguide.
\newblock {\em Phys. Rev. A}, 96:063826, 2017.



\bibitem{Laasko14}
M. Laakso and M. Pletyukhov.
\newblock Scattering of two photons from two distant qubits: exact solution.
\newblock {\em Phys. Rev. Lett.}, 113:183601, 2014.

\bibitem{LK09}
J. S. Li and N. Khaneja.
\newblock Ensemble control of Bloch equations.
\newblock {\em IEEE Trans. Automat. Contr.}, 54:528--536, 2009.

\bibitem{LMS+17}
P. Lodahl, S. Mahmoodian, S. Stobbe, A. Rauschenbeutel, P. Schneeweiss,
  J. Volz, H. Pichler, and P. Zoller.
\newblock Chiral quantum optics.
\newblock {\em Nature}, 541(7638):473--480, 2017.

\bibitem{Nysteen:14}
A. Nysteen, P. T. Kristensen, D. McCutcheon, P. Kaer, and J. M{\o }rk.
\newblock Scattering of two photons on a quantum emitter in a one-dimensional waveguide: Exact dynamics and induced correlations.
\newblock {\em New J. Phys.}, 17:023030, 2015.


\bibitem{Nysteen15}
A. Nysteen, P. T. Kristensen, P. Kaer, and J. M{\o }rk.
\newblock Strong nonlinearity-induced correlations for counterpropagating photons scattering
  on a two-level emitter.
 and induced correlations.
\newblock {\em  Phys. Rev. A}, 91:063823, 2015.


\bibitem{Roulet16}
 A. Roulet, H. N. Le, and V. Scarani.
\newblock  Two photons on an atomic beam splitter: Nonlinear scattering and induced correlations.
\newblock   {\em Phys. Rev. A}, 93: 033838, 2016.

\bibitem{RL00}
R. Loudon.
\newblock {\em The Quantum Theory of Light, 3rd ed}.
\newblock Oxford University Press, Oxford, 2000.

\bibitem{MvH07}
M. Mirrahimi and R. van Handel.
\newblock Stabilizing feedback controls for quantum system.
\newblock {\em SIAM J. Control and Optim.}, 46:445--467, 2007.

\bibitem{Neumeier13}
L. Neumeier, M. Leib, and M. J. Hartmann.
\newblock Single-photon transistor in circuit quantum electrodynamics.
\newblock {\em Phys. Rev. Lett.}, 111:063601, 2013.



\bibitem{NC10} 
M. A. Nielsen  and I. L. Chuang,  
\newblock  {\em Quantum Computation and Information.}
\newblock Cambridge University Press, London, 2010.

\bibitem{NY17}
H. I. Nurdin and N. Yamamoto.
\newblock {\em Linear Dynamical Quantum Systems: Analysis, Synthesis, and
  Control}.
\newblock Springer, 2017.


\bibitem{NJD09}
H. I. Nurdin, M.R. James, and A.C. Doherty.
\newblock Network synthesis of linear dynamical quantum stochastic systems.
\newblock {\em SIAM J. Control and Optim}, 48:2686--2718, 2009.

\bibitem{Pan16}
Y. Pan, D. Dong, and G. Zhang.
\newblock Exact analysis of the response of quantum systems to two-photons
  using a QSDE approach.
\newblock {\em New J. Phys.}, 18(3):033004, 2016.

\bibitem{PZJ15}
Y. Pan, G. Zhang, and M. R. James.
\newblock Analysis and control of quantum finite-level systems driven by
  single-photon input states.
\newblock {\em Automatica}, 69:18--23, 2016.

\bibitem{BQ13}
B. Qi.
\newblock A two-step strategy for stabilizing control of quantum systems with
  uncertainties.
\newblock {\em Automatica}, 49:834--839, 2013.



\bibitem{RWF17}
D. Roy, C. M. Wilson, and O. Firstenberg.
\newblock Colloquium:strongly interacting photons in one-dimensional continuum.
\newblock {\em Rev. Mod. Phys.}, 89:021001, May 2017.

\bibitem{ESS11}
E. Rephaeli, S. E. Kocabaş, and S. Fan.
\newblock Few-photon transport in a waveguide coupled to a pair of colocated two-level atoms.
\newblock {\em Phys. Rev. A}, 84:063832, 2011.

\bibitem{Shen07}
J. T. Shen and S. Fan.
\newblock Strongly correlated two-photon transport in a one-dimensional
  waveguide coupled to a two-level system.
\newblock {\em Phys. Rev. Lett.}, 98:153003, 2007.

\bibitem{SZX16}
H.T. Song, G. Zhang, and Z.  Xi.
\newblock Continuous-mode multiphoton filtering.
\newblock {\em SIAM Journal on Control and Optim.}, 54(3):1602--1632, 2016.



\bibitem{vHSM05}
R. van Handel, J. K. Stockton, and H. Mabuchi.
\newblock Feedback control of quantum state reduction.
\newblock {\em IEEE Trans. Automat. Contr.}, 50:768--780, 2005.

\bibitem{vLFL+13}
A. F. van Look, A. Fedorov, K. Lalumi{\`e}re, B. C. Sanders, A. Blais, and
  A. Wallraff.
\newblock Photon-mediated interactions between distant artificial atoms.
\newblock {\em Science}, 342(6165):1494--1496, 2013.

\bibitem{WM08}
D. F. Walls and G. J Milburn,
\newblock {\em Quantum Optics, 2nd ed.},
\newblock Berlin: Springer, 2008.


\bibitem{WS10}
X. T. Wang and S. G. Schirmer.
\newblock Analysis of Lyapunov method for control of quantum states.
\newblock {\em IEEE Trans. Automat. Contr.}, 55:2259--2270, 2010.

\bibitem{WMS+11}
Y. Wang, J. Min\'a\ifmmode \check{r}\else \v{r}\fi{}, L. Sheridan, and V. Scarani.
\newblock Efficient excitation of a two-level atom by a single photon in a
  propagating mode.
\newblock {\em Phys. Rev. A}, 83:063842, Jun 2011.

\bibitem{WM10}
H. W. Wiseman and G. J. Milburn.
\newblock {\em Quantum Measurement and Control}.
\newblock Cambridge University Press, Cambridge, UK, 2010.


\bibitem{SES13}
S. Xu, E. Rephaeli, and S. Fan.
\newblock Analytic properties of two-photon scattering matrix in integrated quantum systems determined by the cluster decomposition principle.
\newblock {\em Phys. Rev. Lett.}, 111:223602, 2013.

\bibitem{YJ14}
N. Yamamoto and M. R. James.
\newblock Zero-dynamics principle for perfect quantum memory in linear
  networks.
\newblock {\em New J. Phys.}, 16(7):073032, 2014.

\bibitem{Z19}
G. Zhang. 
\newblock Single-photon coherent feedback control and filtering.
\newblock {\em In: Baillieul J., Samad T. (eds) Encyclopedia of Systems and Control.}
\newblock Springer, London, 2020. {\em arXiv:1902.10961v2 [quant-ph]},


\bibitem{ZJ12}
G. Zhang and M. R. James.
\newblock Quantum feedback networks and control: a brief survey.
\newblock {\em Chinese Science Bulletin}, 57(18):2200-2214, 2012.

\bibitem{ZJ13}
G. Zhang and M. R. James.
\newblock On the response of quantum linear systems to single photon input
  fields.
\newblock {\em IEEE Trans. Automat. Contr.}, 58:1221--1235, 2013.

\bibitem{ZGPG18}
G. Zhang, S. Grivopoulos, I. R. Petersen, and J. E. Gough. 
\newblock The Kalman decomposition for linear quantum systems. 
\newblock {\em IEEE Trans. Automat. Contr.}, 63(2):331-346, 2018.


\bibitem{ZB13}
H. Zheng and H. U. Baranger.
\newblock Persistent quantum beats and long-distance entanglement from waveguide-mediated interactions.
\newblock {\em Phys. Rev. Lett.}, 110, 113601, 2013.

\bibitem{ZGB10}
H. Zheng, D. J. Gauthier, and H. U. Baranger.
\newblock QED: Many-body bound-state effects in coherent and Fock-state scattering from a two-level system.
\newblock {\em Phys. Rev. A.}, 82:063816, 2010. 

\bibitem{ZLW+17} 
J. Zhang, Y. X. Liu, R.-B. Wu, K. Jacobs, and F. Nori.
\newblock Quantum feedback: theory, experiments, and applications.
\newblock {\em Physics Reports}, 679: 1-60, 2017.

\end{thebibliography}

\end{document}